 \pdfoutput=1 
\newcommand*{\ATLASLATEXPATH}{latex/}
\documentclass[cernpreprint,texlive=2016,txfonts,UKenglish]{\ATLASLATEXPATH atlasdoc} 

\usepackage[maxbibnames=5]{\ATLASLATEXPATH atlaspackage}
\usepackage{multirow}
\usepackage{\ATLASLATEXPATH atlasbiblatex}

\usepackage{\ATLASLATEXPATH atlascontribute}

\usepackage{\ATLASLATEXPATH atlasphysics}

\graphicspath{{logos/}{figures/}}

\usepackage{IncDiphoton8TeV-defs}

\AtlasJournalRef{Phys. Rev. D 95, 112005 (2017)}
\AtlasDOI{10.1103/PhysRevD.95.112005} 

\AtlasTitle{Measurements of integrated and differential cross sections for isolated photon pair production in $\boldsymbol{pp}$ collisions at $\boldsymbol{\sqrt{s}=8~\TeV}$ with the ATLAS detector}

 \usepackage{authblk}
 \author{The ATLAS Collaboration%
 }

\date{\today}

\AtlasRefCode{STDM-2015-15}

 \PreprintIdNumber{CERN-EP-2017-040} 

 \AtlasJournal{Phys.\ Rev.\ D.}

\AtlasAbstract{%
A measurement of the production cross section for two isolated photons in proton--proton collisions at a center-of-mass energy of
$\sqrt{s}=8~\TeV$ is presented.
The results are based on an integrated luminosity of 20.2~fb$^{-1}$ recorded by the ATLAS detector at the Large Hadron Collider.
The measurement considers photons with pseudorapidities satisfying $|\eta^{\gamma}|<1.37$ or ${1.56<|\eta^{\gamma}|<2.37}$ and transverse energies of respectively $E_{\mathrm{T,1}}^{\gamma}>40~\GeV$ and $E_{\mathrm{T,2}}^{\gamma}>30~\GeV$ for the two leading photons ordered in transverse energy produced in the interaction. The background due to hadronic jets and electrons is subtracted using data-driven techniques. The fiducial cross sections are corrected for detector effects and measured differentially as a function of six kinematic observables. The measured cross section integrated within the fiducial volume is $16.8 \pm 0.8$~pb. The data are compared to fixed-order QCD calculations at next-to-leading-order and next-to-next-to-leading-order accuracy as well as next-to-leading-order computations including resummation of initial-state gluon radiation at next-to-next-to-leading logarithm or matched to a parton shower, with relative uncertainties varying from 5\% to 20\%.
}

\hypersetup{pdftitle={ATLAS document},pdfauthor={The ATLAS Collaboration}}

\begin{document}

\maketitle

\section{Introduction}
\label{sec:introduction}

Diphoton production offers a testing ground for perturbative quantum chromodynamics (pQCD) at hadron colliders, a clean final state for the study of the properties of the Higgs boson and a possible window into new physics phenomena. Cross-section measurements of isolated photon pair production at hadron colliders were performed by the \textsc{D\O} and CDF
collaborations at a center-of-mass energy of $\sqrt{s}=1.96~\TeV$ at the Tevatron proton--antiproton collider~\cite{D0_diphoton,CDF_diphoton} and by the ATLAS and CMS collaborations using ${\sqrt{s}=7~\TeV}$ proton--proton ($pp$) collisions recorded at the Large Hadron Collider (LHC) in 2010 and 2011~\cite{old7tevATLAS,7tevATLAS,Chatrchyan:2011qt,7tevCMS}. Studies of the Higgs-boson's properties in the diphoton decay mode~\cite{Aad:2014eha, Khachatryan:2014ira} and searches for new resonances have also been conducted at the LHC at higher center-of-mass energies~\cite{Aaboud:2016tru, Khachatryan:2016hje}.

The production of pairs of prompt photons, i.e. excluding those originating from interactions with the material in the detector and hadron or $\tau$ decays, in $pp$ collisions can be understood at leading order (LO) via quark--antiquark annihilation ($q\bar{q}$). Contributions from radiative processes and higher-order diagrams involving gluons ($g$) in the initial state are, however, significant. Theoretical predictions, available up to next-to-next-to-leading order (NNLO) in pQCD~\cite{Binoth:1999qq,Catani:2011qz,Campbell:2016yrh} are in good agreement with the total measured diphoton cross sections, while differential distributions of variables such as the diphoton invariant mass (\mgg) show discrepancies~\cite{7tevATLAS}. Theoretical computations making use of QCD resummation techniques~\cite{Ladinsky:1993zn,Resbos:2006PLB,Resbos:2007PRD,Resbos:2007PRDP,Cieri:2015rqa} are also available. These are expected to provide an improved description of the regions of phase space sensitive to multiple soft-gluon emissions such as configurations in which the two photons are close to being back to back in the plane transverse to the original collision.

Significant improvements in the estimation of the backgrounds in the selected sample have been achieved so that the systematic
uncertainties have been reduced by up to a factor of two compared to the previous ATLAS measurement~\cite{7tevATLAS}, despite an increase in the mean number of $pp$ 
interactions per bunch crossing (pileup) from 9.1 at $\sqrt{s} = 7~\TeV$ to 20.7 in the 2012 data sample at $\sqrt{s} = 8~\TeV$. 

In addition to the integrated cross section, differential distributions of several kinematic observables are studied: \mgg, the absolute value of the cosine of the scattering angle with respect to the direction of the proton beams\footnoteATLAS\  expressed as a function of the difference in pseudorapidity between the photons $\left( \acosthetastareta=\tanh\frac{|\Delta\eta_{\gamma\gamma}|}{2}\right)$, the opening angle between the photons in the azimuthal plane (\deltaphi), the diphoton transverse momentum (\ptgg), the transverse component of \ptgg\ with respect to the thrust axis\footnoteThrust\ (\at)~\cite{aTa} and the \phistareta\ variable, defined as ${\phistareta = \tan \left(\frac{\pi - \deltaphi}{2}\right)\sin\theta^{*}_{\eta}}$~\cite{newvarpaper}. Angular variables are typically measured with better resolution than the photon energy. Therefore, a particular reference frame that allows \acosthetastareta\ and $\phistareta$ to be expressed in terms of angular variables only, denoted by the subscript $\eta$ and described in Ref.~\cite{newvarpaper}, is used in order to optimize the resolution. 

The kinematic variables \mgg, \deltaphi~and \ptgg~were studied at $\sqrt{s}=7~\TeV$ in the previous ATLAS publications~\cite{old7tevATLAS,7tevATLAS}. The cosine of the scattering angle with respect to the direction of the proton beams has also been studied, however, in the Collins--Soper frame~\cite{Collins:1977iv}. This is the first time the variables \at~and \phistareta~are measured for a final state other than pairs of charged leptons~\cite{newref1,newref2,Aad:2015auj}. These variables are less sensitive to the energy resolution of the individual photons and therefore are more precisely determined than \ptgg~\cite{aTa,newvarpaper}. Hence, they are ideally suited to probe the region of low \ptgg, in which QCD resummation effects are most significant. Measurements of \ptgg\, \at\ and \phistareta\ for diphoton production (which originates from both the quark--antiquark and gluon--gluon initial states) are important benchmarks to test the description of the low transverse-momentum region by pQCD and complementary to similar measurements performed for $H\rightarrow\gamma\gamma$~\cite{HIGG-2013-10} (in which gluon--gluon initial states dominate) and Drell--Yan (DY)~\cite{newref1,newref2,Aad:2015auj} events (in which quark--antiquark initial states dominate). A good understanding of the low transverse-momentum region in such processes constitutes an important prerequisite for pQCD resummation techniques aiming to describe more complicated processes, e.g. those involving colored final states.

Detailed tests of the dynamics of diphoton production can be performed with the simultaneous study of different differential distributions. Specific regions of the phase space are particularly sensitive to soft gluon emissions, higher-order QCD corrections or non-perturbative effects. Variables such as \mgg\ and |\costhetastareta|\ are also useful in searches for new resonances and measurements of their properties such as their mass and spin. The measurements are compared to recent predictions, which include fixed-order computations up to NNLO in pQCD and also computations combining next-to-leading-order (NLO) matrix elements with a parton shower or with resummation of soft gluons at next-to-next-to-leading-logarithm (NNLL) accuracy.

\section{ATLAS detector}
\label{sec:atlas_detector}

The ATLAS detector~\cite{PERF-2007-01} is a multipurpose detector with a forward-backward symmetric
cylindrical geometry. The most relevant systems for the present measurement are the inner detector (ID), immersed in a $2$\ T
magnetic field produced by a thin superconducting solenoid, and the calorimeters.
The ID consists of fine-granularity pixel and microstrip detectors covering
the pseudorapidity range $|\eta|<2.5$, complemented by a gas-filled straw-tube transition radiation tracker (TRT) that covers the region up to $|\eta| = 2.0$ and provides electron identification capabilities. The electromagnetic (EM) calorimeter is a lead/liquid-argon sampling
calorimeter with accordion geometry. It is divided into a barrel section covering
$|\eta|<1.475$ and two end-cap sections covering $1.375<|\eta|<3.2$. For ${|\eta|<2.5}$, it is divided
into three layers in depth, which are finely segmented in $\eta$ and $\phi$. A thin presampler
layer, covering $|\eta|<1.8$, is used to correct for fluctuations in upstream energy losses.
The hadronic calorimeter in the region $|\eta|<1.7$ uses steel absorbers and scintillator tiles as the
active medium, while copper absorbers with liquid argon are used in the hadronic end-cap sections, which cover the region ${1.5<|\eta|<3.2}$.  A forward calorimeter using copper or
tungsten absorbers with liquid argon completes the calorimeter coverage up to $|\eta|=4.9$. Events are selected using a three-level trigger system. The first level, implemented in custom electronics, reduces the event rate to at most 75~kHz using a subset of detector information~\cite{Aad:2012xs}.
Software algorithms with access to the full detector information are then used in the high-level trigger to yield a recorded event rate of about 400~Hz.

\section{Data sample and event selection}
\label{sec:selection}

The data used in this analysis were recorded using a diphoton trigger with \et\ thresholds of 35~\GeV\ and 25~\GeV\
for the \et-ordered leading and subleading photon candidates, respectively. 
In the high-level trigger, the shapes of the energy depositions in the EM calorimeter are required to match those expected for electromagnetic showers initiated by photons.
The signal efficiency of the trigger~\cite{ATLAS-CONF-2012-048}, estimated using data events recorded with alternative trigger requirements, is $(99.4\pm 0.2)\%$ for events fulfilling the final event selection. Only events acquired in stable beam conditions and passing detector and data-quality requirements are considered. Furthermore, in order to reduce non-collision backgrounds, events are required to have a reconstructed primary vertex with at least three associated tracks with a transverse momentum above $400~\MeV$ and consistent with the average beam spot position. The  effect of this requirement on the signal efficiency is negligible. 
The integrated luminosity of the collected sample is $20.2\pm 0.4$~fb$^{-1}$~\cite{final_lumi_2012}. 

Photon and electron candidates are reconstructed from clusters of energy deposited in the EM calorimeter, and tracks and conversion vertices reconstructed in the ID~\cite{PhotonID:paper}. About 20\% (50\%) of the photon candidates in the barrel (end-cap) are associated with tracks or conversion vertices and are referred to as converted photons in the following. Photons reconstructed within $|\eta| < 2.37$ are retained. Those near the region between the barrel and end-caps ($1.37 < |\eta| < 1.56$) or regions of the calorimeter affected by readout or high-voltage failures are excluded from the analysis.

A dedicated energy calibration~\cite{Aad:2014nim} of the clusters using a multivariate regression algorithm developed and optimized on simulated events is used to determine corrections that account for the energy deposited in front of the calorimeter and outside the cluster, as well as for the variation of the energy response as a function of the impact point on the calorimeter. The intercalibration of the layer energies in the EM calorimeter is evaluated with a sample of $Z$-boson decays to electrons (\Zee) and muons, while the overall energy scale in data and the difference in the effective constant term of the energy resolution between data and simulated events are estimated primarily with a \Zee~sample. Once those corrections are applied, the two photons reconstructed with the highest transverse energies $E_{\mathrm{T,1}}^{\gamma}$ and $E_{\mathrm{T,2}}^{\gamma}$ in each event are retained. Events with $E_{\mathrm{T,1}}^{\gamma}$ and $E_{\mathrm{T,2}}^{\gamma}$ greater than 40~\GeV\ and 30~\GeV, respectively, and angular separation between the photons $\Delta R_{\gamma\gamma} > 0.4$ are selected.

The vertex from which the photon pair originates is selected from the reconstructed collision vertices using a neural-network algorithm~\cite{Aad:2014eha,Aad:2015zhl}. An estimate of the $z$-position of the vertex is provided by the longitudinal and transverse segmentation of the electromagnetic calorimeter, tracks from photon conversions with hits in the silicon detectors, and a constraint from the beam-spot position. The algorithm combines this estimate, including its uncertainty, with additional information from the tracks associated with each reconstructed primary vertex. The efficiency for selecting a reconstructed primary vertex within 0.3~mm of the true primary vertex from which the photon pair originates is about 83\% in simulated events. The trajectory of each photon is then measured by connecting the impact point in the calorimeter with the position of the selected diphoton vertex. 

The dominant background consists of events where one or more hadronic jets contain $\pi^0$ or $\eta$ mesons that carry most of the jet energy and decay to a nearly collinear photon pair. The signal yield extraction, explained in Section~\ref{sec:signal_extraction}, is based on variables that describe the shape of the electromagnetic showers in the calorimeter (\textit{shower shapes})~\cite{PhotonID:paper} and the isolation of the photon candidates from additional activity in the detector.  An initial \textit{loose} selection is derived using only the energy deposited in the hadronic calorimeter and the lateral shower shape in the second layer of the electromagnetic calorimeter, which contains most of the energy. The final \textit{tight} selection applies stringent criteria to these variables and additional requirements on the shower shape in the finely segmented first calorimeter layer. The latter criteria ensure the compatibility of the measured shower profile with that originating from a single photon impacting the calorimeter and are different for converted and unconverted photons. The efficiency of this selection, called identification efficiency in the following, is typically 83–95\% (87–99\%) for unconverted (converted) photons satisfying $30 < E_{\mathrm{T}}^{\gamma} < 100~\GeV$ and isolation criteria similar to the ones defined below.

Additional rejection against jets is obtained by requiring the photon candidates to be isolated using both calorimeter and tracking detector information. The calorimeter isolation variable \etiso\ is defined as the sum of the \et\ of positive-energy topological clusters~\cite{Aad:2016upy} reconstructed in a cone of size $\Delta R = 0.4$ around the photon candidate, excluding the energy deposited in an extended fixed-size cluster centered on the photon cluster. The energy sum is corrected for the portion of the photon's energy deposited outside the extended cluster and contributions from the underlying event and pileup. The latter two corrections are computed simultaneously on an event-by-event basis using an algorithm described in Refs.~\cite{Cacciari:2009dp, STDM-2010-08}. The track isolation variable \ptisol\ is  defined as the scalar sum of the transverse momenta of tracks with $\pt > 1~\GeV$ within a cone of size $\Delta R = 0.2$ around the photon candidate. Tracks associated with a photon conversion and those not originating from the diphoton production vertex are excluded from the sum. Photon candidates are required to have \etiso\ (\ptisol) smaller than 6~\GeV\ (2.6~\GeV). The isolation criteria were re-optimized for the pileup and center-of-mass energy increase in 2012 data. The combined per-event efficiency of the isolation criteria is about 90\% in simulated diphoton events.

A total of 312\,754 events pass the signal selection described above, referred to as inclusive signal region in the following. The largest diphoton invariant mass observed is about 1.7~\TeV.

\section{Monte Carlo simulation}
\label{sec:samples}

Monte Carlo (MC) simulations are used to investigate signal and background properties. Signal and $\gamma$~+~jet background events are generated with both \textsc{Sherpa}~1.4.0~\cite{sherpa} and \textsc{Pythia}~8.165~\cite{pythia}. Drell--Yan ${Z/\gamma^* \rightarrow ee}$ events generated using \textsc{Powheg-box}~\cite{Nason:2004rx,Frixione:2007vw,Alioli:2010xd} combined with \textsc{Pythia} 8.165 are used to study the background arising from electrons reconstructed as photons.

The \textsc{Sherpa} event generator uses the NLO CT10~\cite{CT10} parton distribution functions (PDFs) and matrix elements calculated with up to two final-state partons at LO in pQCD. The matrix elements are merged with the \textsc{Sherpa} parton-shower algorithm~\cite{Schumann:2007mg} following the ME+PS\symbol{64}LO prescription~\cite{Hoeche:2009rj}. The combination of tree-level matrix elements with additional partons and the \textsc{Sherpa} parton-shower effectively allows a simulation of all photon emissions~\cite{Siegert:2016bre}. A modeling of the hadronization and underlying event is also included such that a realistic simulation of the entire final state is obtained. In order to avoid collinear divergences in the generation of the $2\rightarrow 3$ and $2\rightarrow 4$ processes, a minimum angular separation between photons and additional partons of $\Delta R>0.3$ is set at the generator level. 

The \textsc{Pythia} signal sample takes into account the $q\bar{q}\rightarrow\gamma\gamma$ and $gg\rightarrow\gamma\gamma$ matrix elements at LO, while the $\gamma$ + jet sample is produced from $qg \rightarrow q\gamma$, $\bar{q}g \rightarrow \bar{q}\gamma$ and $q\bar{q} \rightarrow g\gamma$ at LO. Photons originating from the hadronization of colored partons are taken into account by considering LO $\gamma$+jet and dijet diagrams with initial- and final-state radiation. All matrix elements are interfaced to the LO CTEQ6L1~\cite{cteq6} PDFs. The \textsc{Pythia} event generator includes a parton-shower algorithm and a model for the hadronization process and the underlying event. 

The \textsc{Powheg-box} DY sample uses matrix elements at NLO in pQCD and the CT10 PDFs with \textsc{Pythia} 8.165 to model parton showering, hadronization and the underlying event. The final-state photon radiation is simulated using \textsc{Photos}~\cite{Golonka:2005pn}.

The \textsc{Pythia} event-generator parameters are set according to the ATLAS AU2 tune~\cite{ATL-PHYS-PUB-2012-003}, while the \textsc{Sherpa} parameters are the default parameters recommended by \textsc{Sherpa} authors. The generated events are passed through a full detector simulation~\cite{Aad:2010ah} based on \textsc{Geant4}~\cite{geant}. Pileup from additional $pp$ collisions in the same and neighboring bunch crossings is simulated by overlaying each MC event with a variable number of simulated inelastic $pp$ collisions generated using \textsc{Pythia} 8.165. The MC events are weighted to reproduce the observed distribution of the average number of interactions per bunch crossing and the size of the luminous region along the beam axis.

Photon and electron energies in the simulation are smeared by an additional effective constant term in order to match the width of the $Z$-boson resonance to that observed in data~\cite{Aad:2014nim}. The probability for a genuine electron to be wrongly reconstructed as a photon (electron-to-photon fake rate) in the DY sample is corrected to account for extra inefficiencies in a few modules of the inner detector as observed in \Zee~events in data. The photon shower shapes and the calorimeter isolation variables in the MC simulation are corrected for the small observed differences in their average values between data and simulation in photon-enriched control samples and \Zee\ decays. Residual differences between data and MC efficiencies of the identification and calorimeter isolation criteria are typically less than 2\% each per photon and are corrected for by using factors~\cite{PhotonID:paper} that deviate from one at most by 6\% per event, in rare cases. The track isolation efficiency is calculated from the ratio of the diphoton yields obtained with and without the application of the track isolation criterion, using the same methods used in the signal extraction, described below. The mean of the results obtained with the different methods in each bin is used and is about 94\%. The effective efficiency corrections for simulated diphoton events are typically 2\% per event.

\section{Signal yield extraction}
\label{sec:signal_extraction}

Two data-driven methods, similar to the ones described in Refs.~\cite{7tevATLAS,old7tevATLAS}, are employed in the signal yield extraction and give consistent results: a two-dimensional template fit method and a matrix method. Both methods determine the sample composition by relying on the identification and isolation discriminants to define jet-enriched data control regions with small contamination from photons. The methods are validated using pseudo-data generated with known signal and background composition by combining events from MC simulation and the data control regions. A \emph{non-tight} selection is constructed by inverting some of the tight identification requirements computed using the energy deposits in cells of the finely segmented first calorimeter layer~\cite{Aad:2010sp}. The isolation energy around the photon candidate, computed in a much wider region, is to a large extent independent of these criteria. Data control regions where both photon candidates satisfy the non-tight selection or either the subleading or the leading candidate satisfies the non-tight selection while the other candidate satisfies the signal selection are used as background control samples to estimate the number of jet + jet (\evtjetjet), $\gamma$ + jet (\evtgamjet) and jet + $\gamma$ (\evtjetgam) events in the selected signal sample,\footnote{In the $\gamma$ + jet (jet + $\gamma$) component, the subleading (leading) candidate is assumed to be a jet misidentified as a photon. In the jet + jet component, both candidates are assumed to be jets misidentified as photons.} respectively. Another background component, typically one order of magnitude smaller, arises from isolated electrons reconstructed as photon candidates. This component is dominated by ${Z\rightarrow ee}$ decays and thus located in terms of invariant mass near the $Z$-boson mass ($80<\mgg<100~\GeV$). Data control regions where one of the candidates is reconstructed as an electron and the other one as a photon are used to study this background. The contribution from other background sources is found to be negligible.

\subsection{The two-dimensional template fit method}
\label{2dfit}

The template fit method consists of an extended maximum-likelihood 
fit to the two-dimensional distribution of the calorimeter isolation variables (\eiso{1}, \eiso{2}) of events passing the signal selection. The yields associated with five components are extracted simultaneously: $\gamma\gamma$, \evtgamjet, \evtjetgam, \evtjetjet\ and dielectrons ($ee$). The dielectron yield is constrained to the value predicted by MC simulation within the uncertainties obtained from the matrix method described in Section~\ref{4x4}. The fit is performed in the inclusive signal region and in each bin of the variables studied. Independent templates are used in each bin in order to account for the dependence of the isolation distributions on the event kinematics, while in Ref.~\cite{7tevATLAS} this effect is considered as a systematic uncertainty. The extraction of the templates is described below. The variation in the distributions for dielectron events in the different kinematic bins do not have a significant impact on the extracted yields and therefore a single template, corresponding to the inclusive selection, is used.

\begin{figure}[t]
\centering
\subfloat[]{\includegraphics[width=0.5\textwidth]{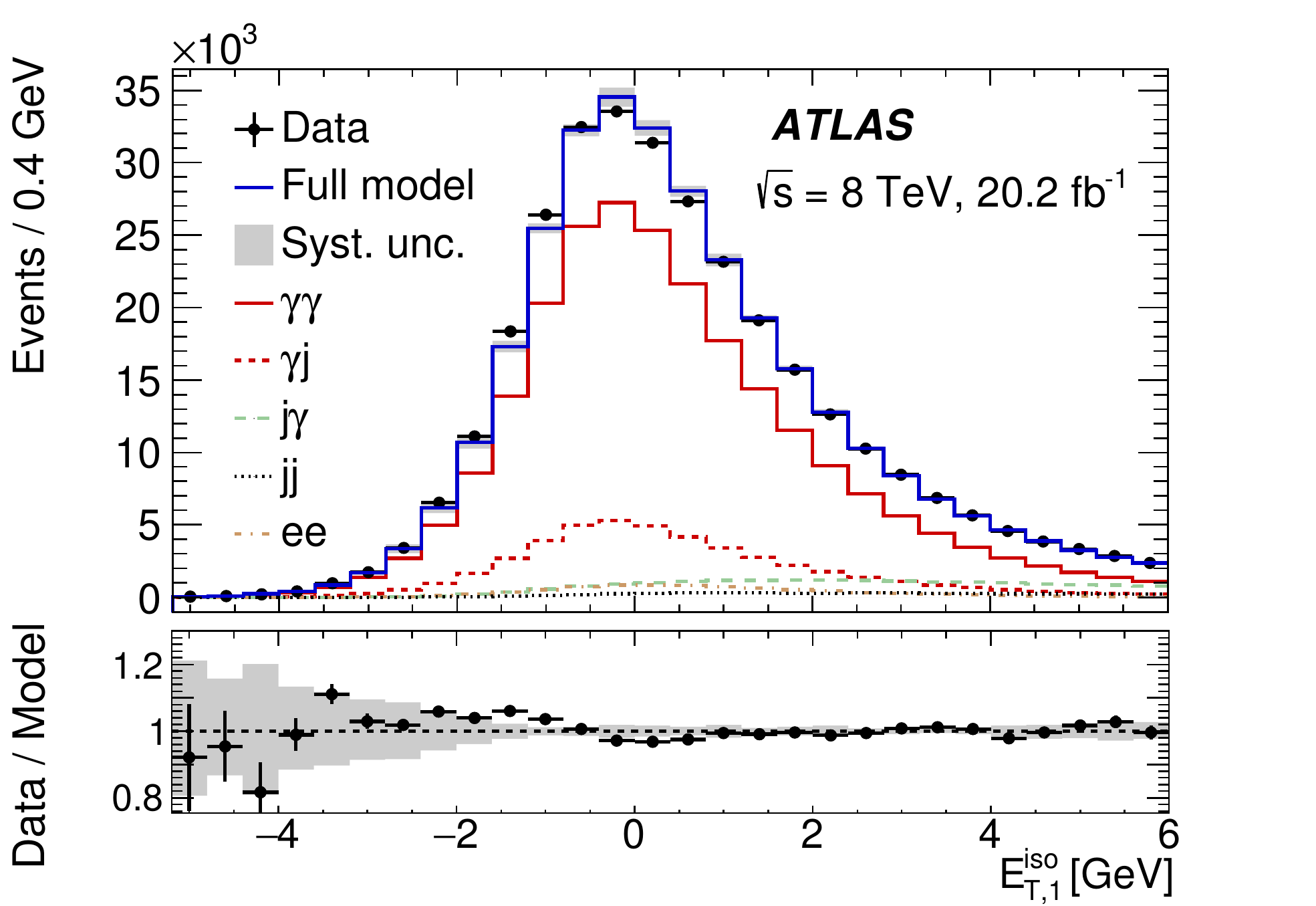}}
\subfloat[]{\includegraphics[width=0.5\textwidth]{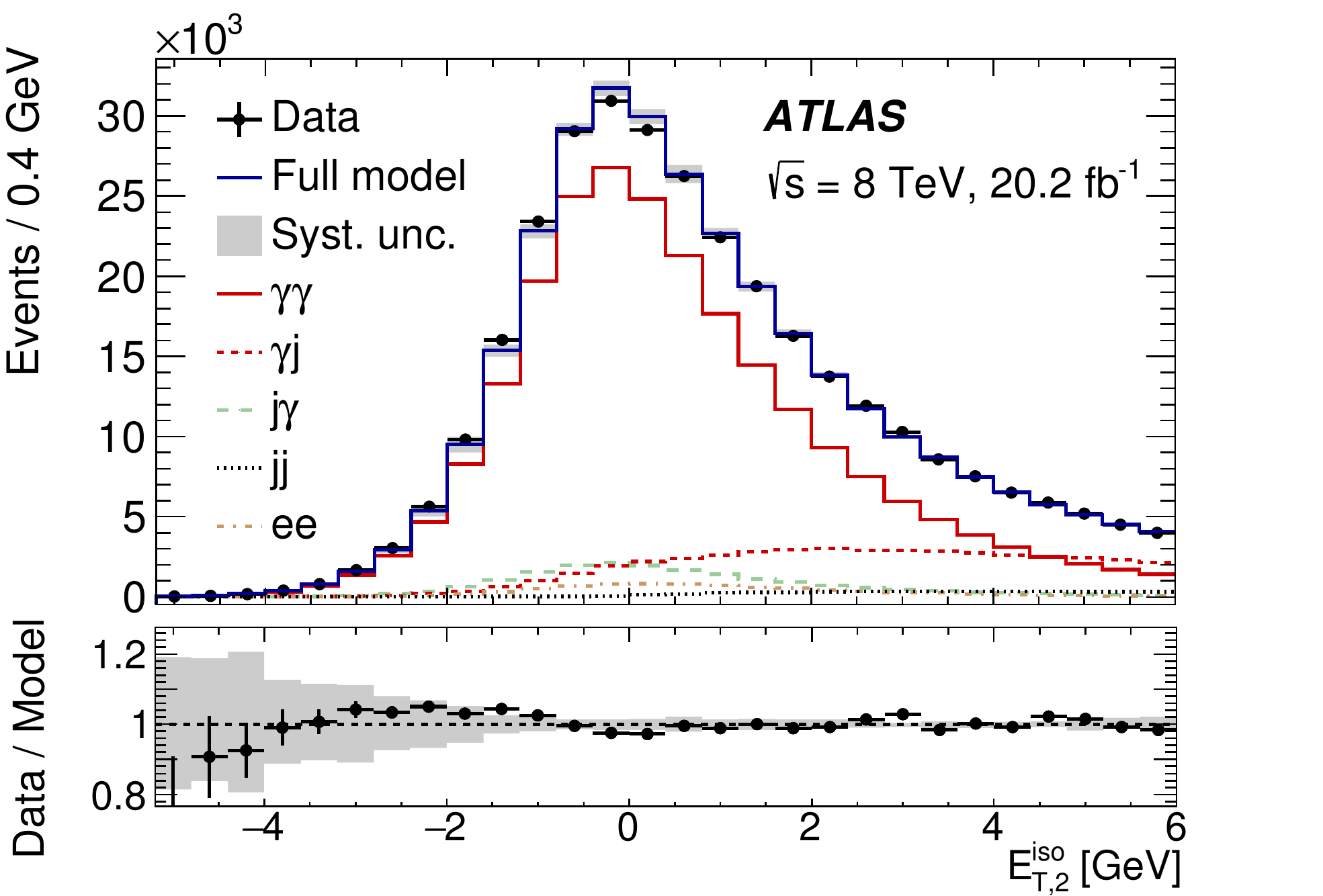}}
\caption{Distributions of the calorimeter transverse isolation energy for the (a) leading and (b) subleading photon candidates. Also shown are the projections of the signal and various background components used in the two-dimensional template fit and the full model, corresponding to their sum, after fit. At the bottom of each plot, the ratio of the data to the model after the fit is shown. A gray uncertainty band on the full model, including the systematic uncertainties discussed in the text, is shown in each case whereas the uncertainty in data is statistical only.}

\label{fig:2dfit_inclusive}
\end{figure}

The templates for diphoton events are obtained from the \textsc{Sherpa} diphoton sample, including the data-driven corrections described in Section~\ref{sec:samples}. The templates associated with the \evtgamjet\ and \evtjetgam\ processes are defined by a product of the one-dimensional distributions associated with the photon candidate in the \textsc{Sherpa} diphoton sample and the jet candidate in the corresponding jet-enriched data control regions, normalized such that the final templates correspond to probability density functions. The templates associated with jj events are taken from the control region where both candidates satisfy the non-tight selection. Neglecting the correlations between the transverse isolation energies of the two candidates in \evtgamgam\ and \evtjetjet\ events would bias the extracted diphoton yields by a few percent, as indicated by tests with pseudo-data. Therefore, full two-dimensional distributions are used to model those components while for the other components the products of one-dimensional distributions associated with each of the candidates are used. The contamination of events containing real photons, relevant for \evtgamjet, \evtjetgam\ and \evtjetjet, is subtracted using the signal MC sample. The signal contamination is about 12\%, 17\% and 2\% for \evtgamjet, \evtjetgam\ and \evtjetjet, respectively. The templates are defined as either binned distributions or smooth distributions via kernel density estimators~\cite{Cranmer:2000du}, the latter being used if there are few events. The signal purity of events passing the inclusive selection is about 75\% and ranges typically from 60\% to 98\% across the bins of the various observables. Figure~\ref{fig:2dfit_inclusive} shows the distributions of \etiso\ for the two photon candidates in the inclusive signal region, together with the projections of the five two-dimensional fit components after the maximization of the likelihood.

Systematic uncertainties originating from the modeling of the calorimeter isolation variables of photons and jets are considered. The extracted diphoton yield changes by $\pm 2.2\%$ when constructing the photon templates using the diphoton events generated by \textsc{Pythia} instead of \textsc{Sherpa}, due to the impact on the isolation distribution of the different treatment of additional partons and photon radiation, as described in Section~\ref{sec:samples}. Varying the data-driven corrections to the calorimeter isolation variables of photons within their uncertainties results in $\pm 1.9\%$ variations in the signal yields. Changing the set of identification requirements that are inverted to construct the non-tight definition impacts the jet templates due to  possible correlations between the identification and isolation criteria. The result is a $^{+1.5}_{-1.7}\%$ variation of the inclusive diphoton yield. Uncertainties due to the photon identification criteria impact the estimated contamination of photon events in the jet-enriched control regions. The corresponding uncertainty in the signal yield is $\pm 0.9$\%. The uncertainties in the estimated number of events containing electrons result in a $-1.0$\% change in the signal yield.  Other sources of uncertainty, related to the modeling of the pileup and fit method were estimated and found to have a small impact on the diphoton yield. The total uncertainty in the inclusive signal yield, obtained by summing the contributions in quadrature, is $^{+3.5}_{-3.7}$\%. Compared to Ref.~\cite{7tevATLAS}, the reduction of the uncertainty 
is achieved due to the higher signal purity provided by the re-optimized isolation requirements and the restriction of the transverse energy of the photons, which is $\et > 40~\GeV$ and 30~\GeV\ compared to 25~\GeV\ and 22~\GeV\ previously. The fit method is also improved, including the better modeling of the calorimeter isolation distributions due to the data-driven corrections, the use of independent templates in each bin and the inclusion of the electron component.

The total systematic uncertainty in the differential measurements ranges typically from 3\% to 10\%. The largest values are observed in regions where the contributions from radiative processes are expected to be large (e.g. for large values of \phistareta\ or small values of \deltaphi) and less well modeled by the MC simulation and with lower signal purity such that the jet control-region definition has a more significant impact on the results. 

\subsection{The matrix method}
\label{4x4}

In the matrix method, the sample composition is determined by solving a linear system of equations $\left(N_{PP}, N_{PF}, N_{FP}, N_{FF}\right)^T = E \cdot \left( N_{\evtgamgam}, N_{\evtgamjet}, N_{\evtjetgam}, N_{\evtjetjet} \right)^T$ that relates the populations associated with each process and the numbers of events in which the candidates pass ($P$) or fail ($F$) the isolation requirements. The matrix $E$ contains the probabilities of each process to produce a certain final state and is constructed using the efficiency of the isolation criteria for photons and jets. The efficiencies are extracted from collision data over a wide isolation energy range, relying on the signal MC sample to estimate the contamination of events containing real photons in the data control regions. The efficiencies are derived in bins of $E_{\mathrm{T}}$, $|\eta|$ and of the corresponding observable for differential measurements. In Ref.~\cite{old7tevATLAS}, only the variations with $E_{\mathrm{T}}$\ and $|\eta|$ are considered. 
The inclusive diphoton yield is obtained from the sum of the yields in each bin of \mgg. 

The electron contamination is determined from \Zee\ decays in data using a method similar to the one described in Ref.~\cite{old7tevATLAS}. The electron-to-photon fake rate ($f_{e \rightarrow \gamma}$) is estimated separately for the leading and subleading candidate from the ratios of events $N_{\gamma e} / N_{ee}$ and $N_{e\gamma} / N_{ee}$ around the $Z$-boson mass in the invariant-mass range [80~\GeV, 100~\GeV], after subtracting the contribution from misidentified jets using the invariant-mass sidebands. The $f_{e \rightarrow \gamma}$ value is estimated to be about 4\% in the calorimeter barrel and up to about 10\% in the calorimeter end-caps, in agreement with the \Zee~MC sample. The probability for photons to be reconstructed as electrons, taken from MC simulation, is less than 1\%. The $\gamma\gamma$, $\gamma e$, $e\gamma$ and $ee$ yields in the selected signal sample are then obtained by solving a linear system of equations relating the true and reconstructed event yields, similarly to the jet-background estimation. In the inclusive signal region, the sum of the contributions of the ${e\gamma}$ and ${\gamma e}$ background is estimated to be about 10\% of the $ee$ background yield. It is included in the estimation of the total electron-background yield for the matrix method and as a systematic uncertainty for the template fit, as described in Section~\ref{2dfit}.

The dominant source of systematic uncertainty affecting the signal yield estimated with the matrix method arises from the definition of jet-enriched control regions ($^{+3.3}_{-2.6}$\%), followed by the contamination of those regions by true photon events ($\pm 2.4\%$), estimated by comparing the predictions from \textsc{Sherpa} and \textsc{Pythia}. Both uncertainties are slightly larger than for the template fit method due to the use of additional control regions with higher isolation energies in the determination of the isolation efficiencies. The uncertainties in the signal yield arising from events containing electrons, mostly related to the jet background contamination in the electron control samples, are $\pm 0.6\%$. The total systematic uncertainty in the diphoton yield is $^{+4.2}_{-3.7}$\%.

\subsection{Sample composition}
\label{sec:final_bkg}

\begin{figure}[htbp]
\includegraphics[width=0.5\textwidth]{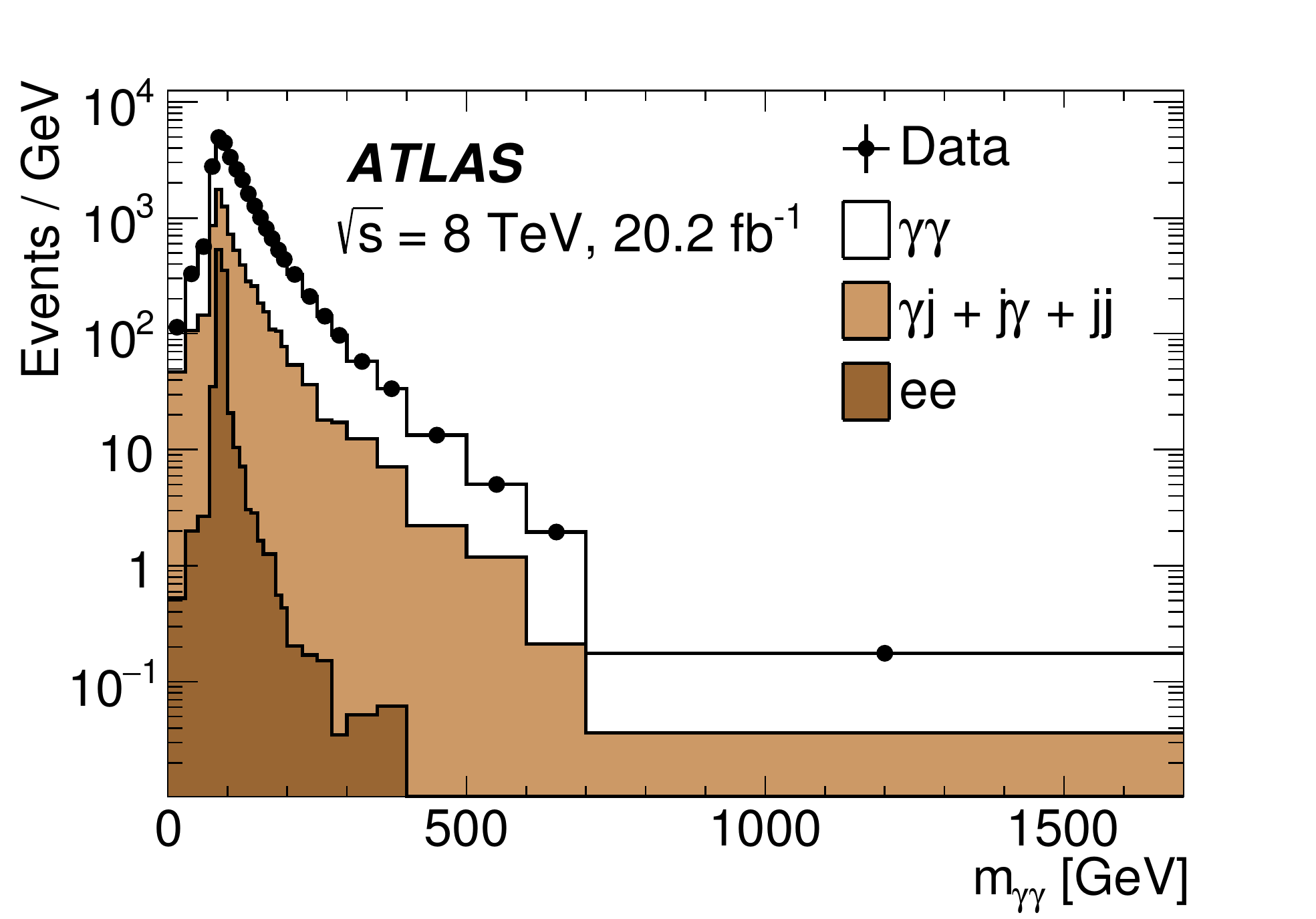}
\includegraphics[width=0.5\textwidth]{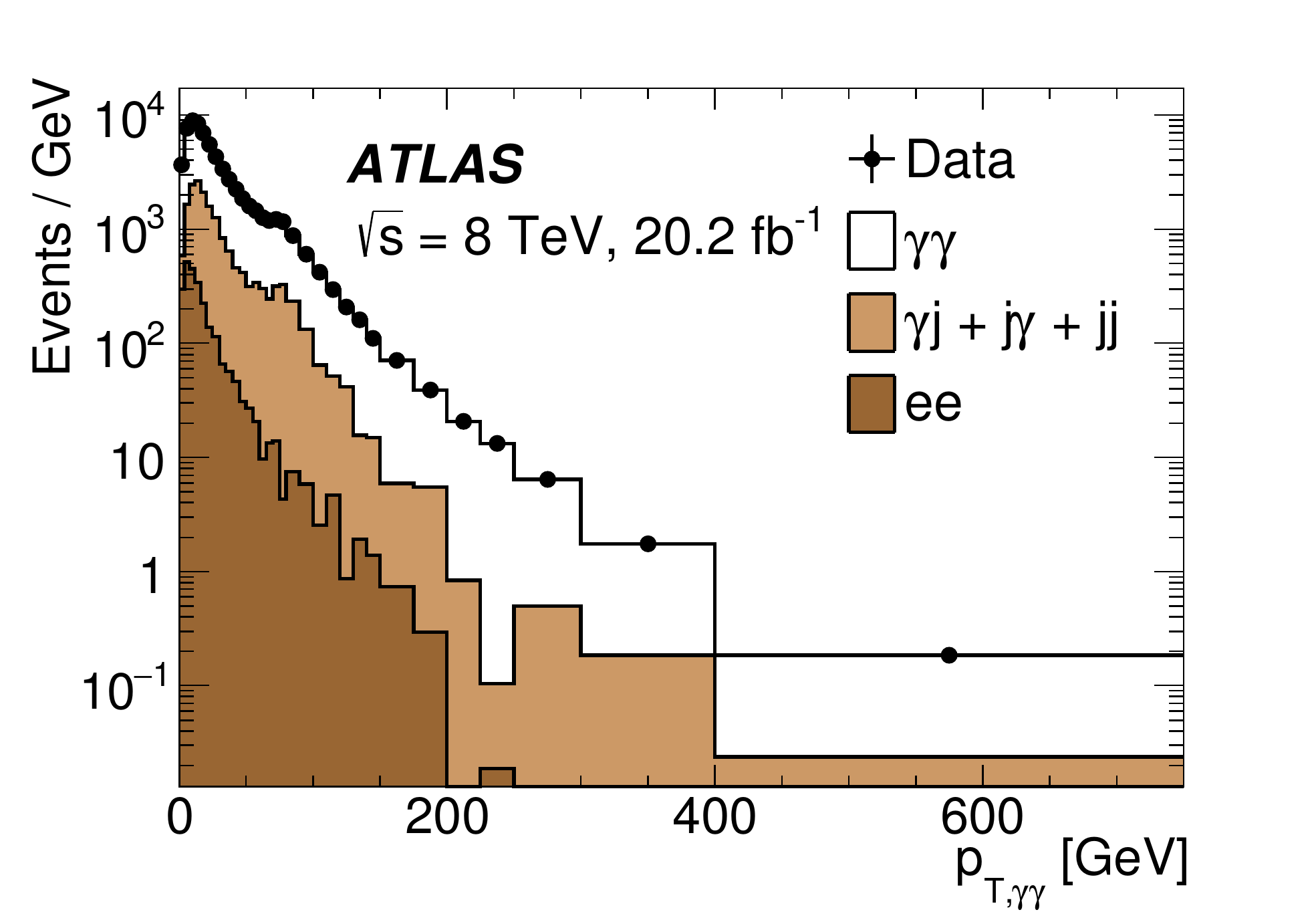}
\includegraphics[width=0.5\textwidth]{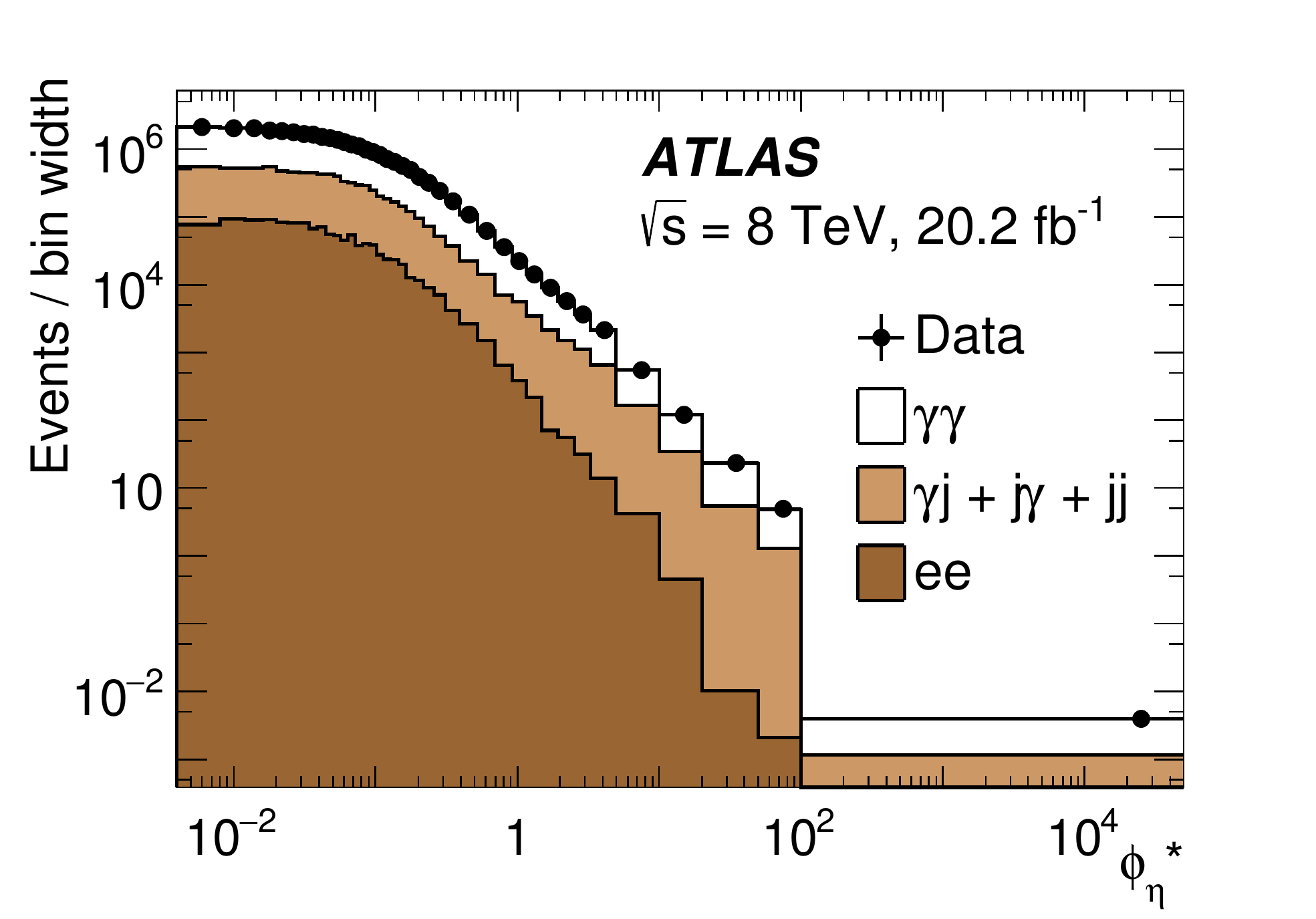}
\includegraphics[width=0.5\textwidth]{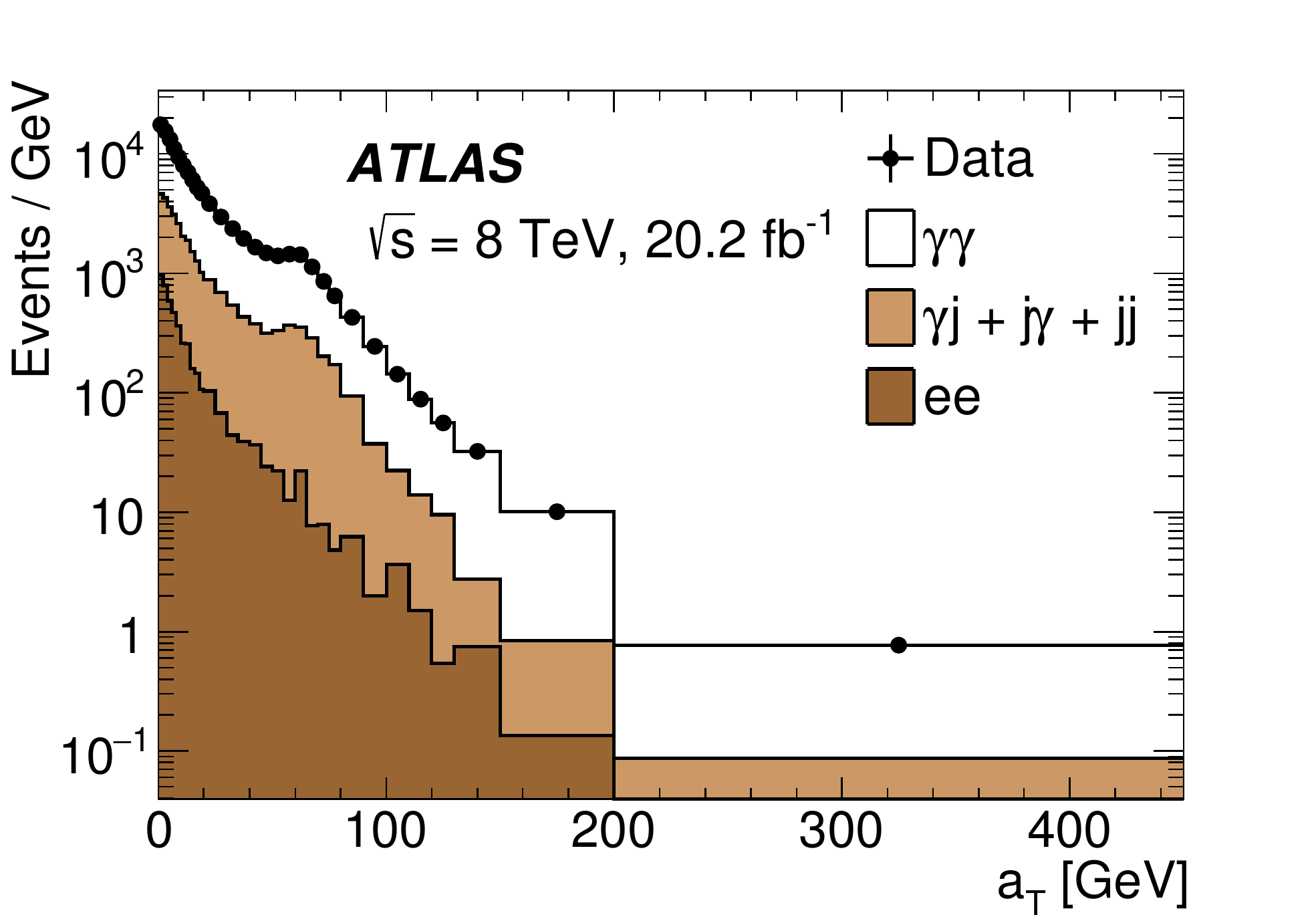}
\includegraphics[width=0.5\textwidth]{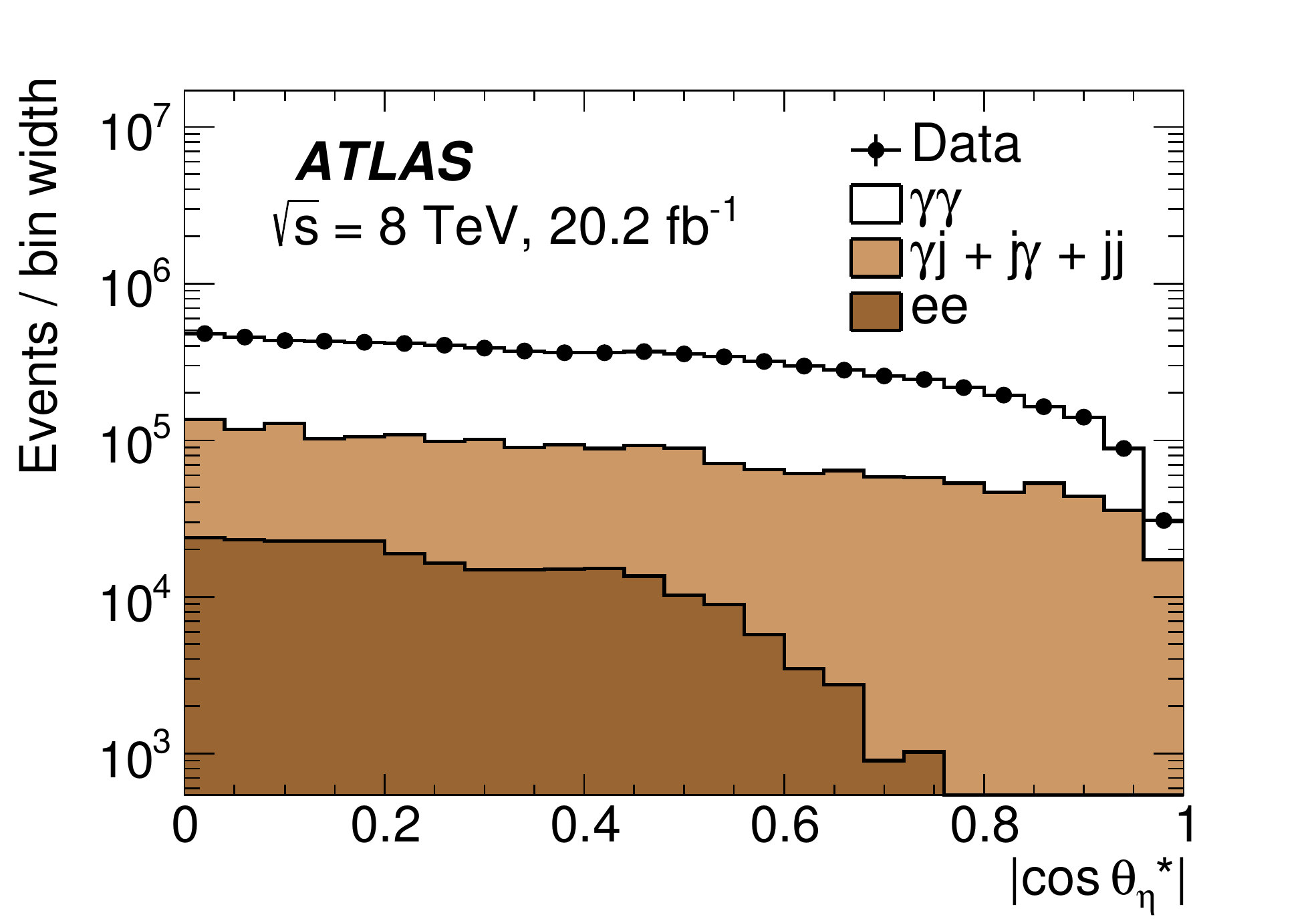}
\includegraphics[width=0.5\textwidth]{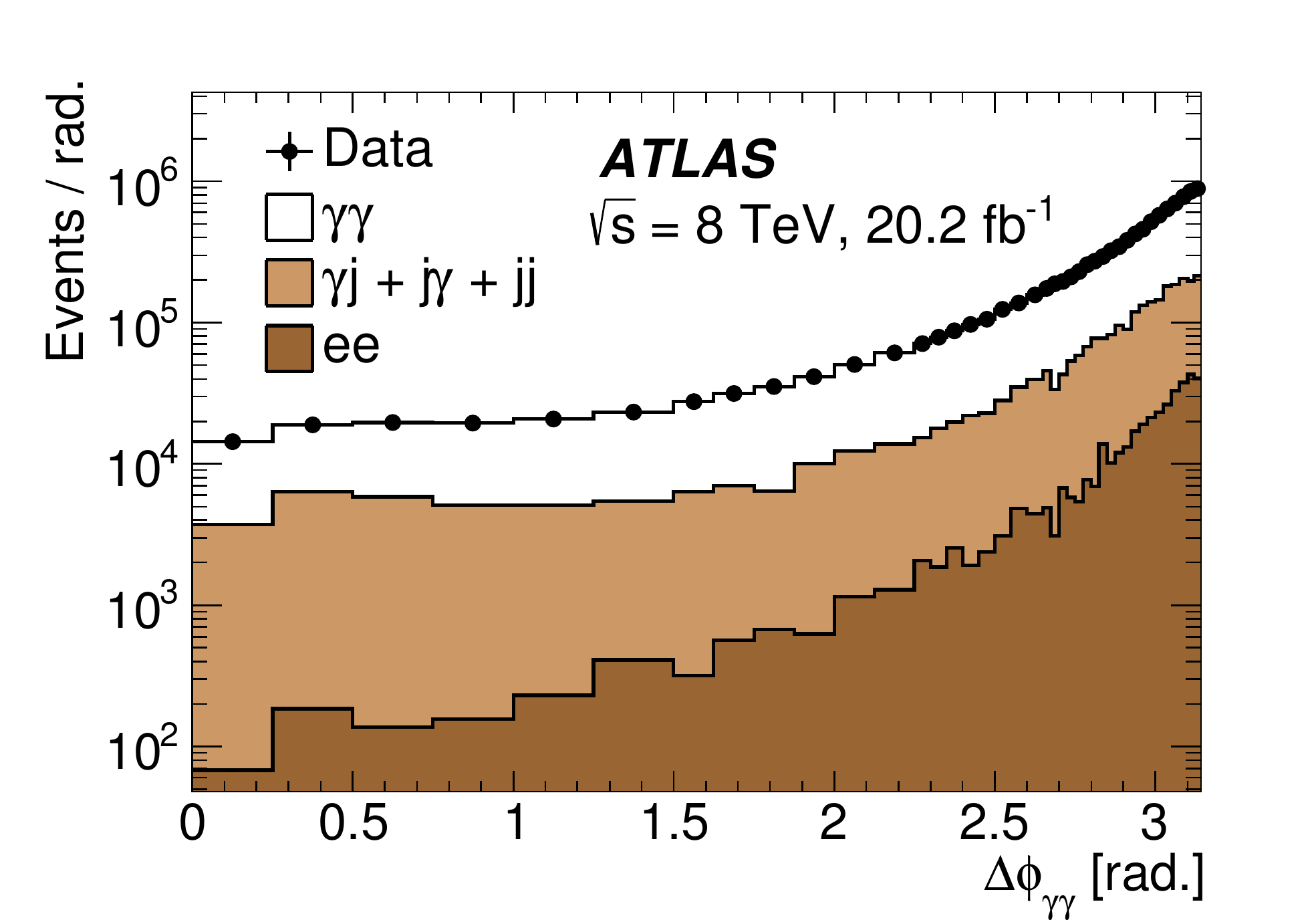}
\caption{Distributions of the reconstructed diphoton observables, together with the yields associated with the various components estimated using the two-dimensional template fit method.}
\label{fig:decomposition}
\end{figure}

The estimated sample composition of the events fulfilling the inclusive selection is shown in Table~\ref{tab:final_inclusive} with associated uncertainties. The composition in the different bins for which the differential cross sections are measured is shown in Figure~\ref{fig:decomposition}. The signal purity increases relative to that of the inclusive selection by 5--15\% (in absolute) in regions dominated by candidates with higher transverse momenta and smaller pseudorapidities, typically towards larger values of \mgg, \ptgg\ and \at. The variations of the purity as a function of \acosthetastareta, \phistareta\ and \deltaphi\ are usually within 5\%.
The template fit method and the matrix method are based on the same isolation and identification criteria to disentangle signal and background and therefore cannot be combined easily.
The main correlations arise from the use of the same signal selection and overlapping background control regions, resulting in a total correlation of about 25\% between the signal yield estimates. Taking into account the correlated and uncorrelated uncertainties, the signal yields returned by the two methods are found to be compatible within one standard deviation for the inclusive selection. The magnitudes of the associated uncertainties are also comparable. The cross-section measurements presented in the following section are derived using the two-dimensional template fit method.

\begin{table}[htb]
  \caption{Estimated sample composition in the inclusive signal region using the two-dimensional template fit and the matrix method. Both the statistical and total systematic uncertainties are listed.}
  \begin{center}
      \begin{tabular}{|c|ll|}
     \hline
 \multirow{2}{*}{Process} & \multicolumn{2}{c|}{Event fraction [\%]} \\
 & Two-dimensional template fit & \multicolumn{1}{c|}{Matrix method} \\
  \hline
 & & \\
${\gamma\gamma}$         &  75.3 $\pm$ 0.3 (stat) $^{+2.6}_{-2.8}$ (syst)  &  73.9 $\pm$ 0.3 (stat) $^{+3.1}_{-2.7}$ (syst)\\
 & & \\
${\gamma\text{j}}$       &  14.5 $\pm$ 0.2 (stat) $^{+2.7}_{-2.8}$ (syst) &  14.4 $\pm$ 0.2 (stat) $^{+2.0}_{-2.4}$ (syst)\\
 & & \\
${\text{j}\gamma }$      &  ~~6.0 $\pm$ 0.2 (stat) $^{+1.4}_{-1.5}$ (syst)  &  ~~5.8  $\pm$ 0.1 (stat) $\pm$ 0.6 (syst)\\
 & & \\
${\text{jj}}$            &  ~~1.6 $\pm$ 0.2 (stat) $^{+0.9}_{-0.4}$ (syst) &  ~~2.4  $\pm$ 0.1 (stat) $^{+0.6}_{-0.5}$ (syst)\\
 & & \\
${ee}$                   &  ~~2.6 $\pm$ 0.2 (stat) $^{+0.9}_{-0.4}$ (syst) &  ~~3.5  $\pm$ 0.1 (stat) $\pm$ 0.4 (syst) \\
 & & \\
 \hline
   \end{tabular}
  \end{center}
  \label{tab:final_inclusive}
\end{table}

\section{Cross-section measurements}
\label{sec:xsection}

The integrated and differential cross sections as functions of the various observables are measured in a fiducial region defined at particle level to closely follow the criteria used in the event selection. The same requirements on the photon kinematics are applied, i.e. $E_{\mathrm{T,1}}^{\gamma}$ and $E_{\mathrm{T,2}}^{\gamma}$ greater than 40~\GeV\ and 30~\GeV, respectively, $|\eta| < 2.37$ excluding $1.37 < |\eta| < 1.56$, and angular separation $\Delta R_{\gamma\gamma} > 0.4$. The photons must not come from hadron or $\tau$ decays. The transverse isolation energy of each photon at particle level, \etisopart, is computed from the sum of the transverse momenta of all generated particles in a cone of $\Delta R = 0.4$ around the photon, with the exception of muons, neutrinos and particles from pileup interactions. A correction for the underlying event is performed using the same techniques applied to the reconstructed calorimeter isolation but based on the four-momenta of the generated particles. The effect of the experimental isolation requirement used in the data is close to the particle-level requirement of $\etisopart < 11~\GeV$, included in the fiducial region definition.

The estimated numbers of diphoton events in each bin are corrected for detector resolution, reconstruction and selection efficiencies using an iterative Bayesian unfolding method~\cite{unfolding,Adye:2011gm}. Simulated diphoton events from the \textsc{Sherpa} sample are used to construct a response matrix that accounts for inefficiencies and bin-to-bin migration effects between the reconstructed and particle-level distributions. The bin purity, defined from the MC signal samples as the fraction of reconstructed events generated in the same bin, is found to be about 70\% or higher in all bins. After five iterations the results are stable within 0.1\%. 

The contribution from the Higgs-boson production and subsequent decay into two photons to the measured cross sections is expected to be at the few-percent level around $120 < \mgg < 130~\GeV$ and negligible elsewhere. Although this process is neglected in the event generation, no correction is applied given its small magnitude.

The cross section in each bin is computed from the corrected number of diphoton events in the bin divided by the product of the integrated luminosity~\cite{final_lumi_2012} and the trigger efficiency. Tabulated values of all measured cross sections are available in the Durham HEP database~\cite{hepdata}. The measured fiducial cross section is:

\begin{linenomath}
\begin{equation}
\sigma_{\mathrm{tot}}^{\mathrm{fid.}} = 16.8 \pm 0.1~\mathrm{(stat)} \pm 0.7~\mathrm{(syst)} \pm 0.3~\mathrm{(lumi)}~\mathrm{pb} = 16.8 \pm 0.8~\mathrm{pb}.
\label{eq:xsec_tot}
\end{equation}
\end{linenomath}

The uncertainties represent the statistical (stat), the systematic (syst) and the luminosity (lumi) uncertainties. The systematic component receives contributions from the uncertainties in the signal yield extraction, signal selection efficiency, including the trigger efficiency (see Section~\ref{sec:selection}), and the unfolding of the detector resolution. The uncertainty in the photon identification efficiency ranges between 0.2\% and 2.0\% per photon, depending on $\et$, $\eta$ and on whether the photon is reconstructed as unconverted or converted~\cite{PhotonID:paper}. The uncertainties are derived from studies with a pure sample of photons from $Z \rightarrow \ell\ell\gamma$ decays (where $\ell$ is an electron or muon), \Zee\ decays by extrapolating the properties of electron showers to photon showers using MC events and isolated photon candidates from prompt $\gamma$ + jet production. The resulting uncertainty in the integrated diphoton cross section, including both the effects on the signal yield extraction and signal selection efficiency, is 2.5\%. The uncertainty in the modeling of the calorimeter isolation variable is 2.0\% and is estimated by varying the data-driven corrections to the MC events within their uncertainties. The systematic uncertainty in the track isolation efficiency is taken as the largest effect among the difference between the efficiencies estimated with the template fit method and the matrix method, and the variation between the efficiencies predicted from simulated events with \textsc{Sherpa} and \textsc{Pythia}. The impact on the integrated cross section is 1.5\%. The uncertainty in the corrections applied during the unfolding procedure is estimated by using diphoton events from \textsc{Pythia} instead of \textsc{Sherpa} to construct the response matrices. The change in these corrections partly compensates for the different signal yields extracted with the two event generators. The combined uncertainty due to the choice of MC event generator is 1.1\%.

The impact of each source of uncertainty is estimated by repeating the signal yield extraction and cross-section measurement with the corresponding changes in the calorimeter isolation templates and unfolding inputs. The total uncertainty is obtained by summing all contributions in quadrature. The main sources of systematic uncertainty and their impact on the integrated cross section are listed in Table~\ref{tab:unc_final}. The uncertainties in the differential cross sections are of the same order. However, the measured cross sections in regions where the contributions from radiative processes are expected to be large (e.g. for large values of \phistareta\ or small values of \deltaphi) are more affected by the choice of MC event generator and the uncertainty in the track isolation efficiency. The variations associated with the former (latter) range between 1\% and 10\% (1\% and 7\%) across the various bins.
Uncertainties in photon energy scale (0.2--0.6\% per photon) and resolution (${\sim}10\%$ per photon), estimated mostly from \Zee\ decays~\cite{Aad:2014nim}, give small contributions to the total uncertainty in the cross section, reaching as high as 4\% at large values of \mgg, \ptgg\ and \at, where the precision in the cross section is limited by statistical uncertainties. Other sources of uncertainty such as those related to the choice of primary vertex, the impact of the modeling of the material upstream of the calorimeter on the photon reconstruction efficiency or the main detector alignment effects like a dilation of the calorimeter barrel by a few per mille or displacements of the calorimeter end-caps by few millimeters have a negligible impact on the measured cross sections.

\begin{table}[htbp]
\setlength\extrarowheight{3pt}
\caption{Main sources of systematic uncertainty and their impact on the integrated fiducial cross section ($\sigma_{\mathrm{tot}}^{\mathrm{fid.}}$).\bigskip}
\centering
\small
\begin{tabular}{|lr|}
\hline
Source of uncertainty & Impact on $\sigma_{\mathrm{tot}}^{\mathrm{fid.}}$ [\%] \\[3.5pt]
\hline
Photon identification efficiency  &  $\pm 2.5$ \\
Modeling of calorimeter isolation  & $\pm2.0$ \\
Luminosity        & $\pm 1.9$  \\
Control-region definition         & $^{+1.5}_{-1.7}$  \\
Track isolation efficiency    & $\pm1.5$ \\
Choice of MC event generator  & $\pm 1.1$  \\
Other sources combined & $^{+0.8}_{-1.3}$ \\[3pt]
\hline
Total                           & $^{+4.5}_{-4.7}$ \\[3.5pt]
\hline
\end{tabular}
\label{tab:unc_final}
\end{table}

\section{Comparison with theoretical predictions}
\label{sec:theory}

The measured integrated fiducial and differential cross sections are compared to fixed-order predictions at NLO and NNLO accuracies in pQCD. They are also compared to  computations combining NLO matrix elements and resummation of initial-state gluon radiation to NNLL or matched to a parton shower.

A fixed-order NLO calculation is implemented in \textsc{Diphox}~\cite{Binoth:1999qq}, including direct and fragmentation contributions as well as the first order of the $gg\rightarrow\gamma\gamma$ component via a quark loop. The \textsc{Resbos} NLO event generator~\cite{Resbos:2006PLB,Resbos:2007PRD,Resbos:2007PRDP} adopts a simplified approach to treat the fragmentation contributions based on a combination of a sharp cutoff on the transverse isolation energy of the final-state quark or gluon and a smooth cone isolation~\cite{Frixione:1998jh}, while the first corrections to the $gg$-initiated process and resummation of initial-state gluon radiation to NNLL accuracy are included. The \textsc{2$\gamma$NNLO}~\cite{Catani:2011qz} program includes the direct part of diphoton production at parton level with NNLO pQCD accuracy but no contributions from fragmentation photons. For the \textsc{Diphox} and \textsc{Resbos} predictions, the transverse energy of partons within a cone of $\Delta R = 0.4$ around the photons is required to be below 11~\GeV. A smooth isolation criterion~\cite{Frixione:1998jh}, described in Refs.~\cite{Andersen:2014efa, Cieri:2015wwa}, is adopted to obtain infrared-safe predictions for \textsc{2$\gamma$NNLO} using a maximum transverse isolation energy of 11~\GeV\ and a cone of $\Delta R = 0.4$.

The CT10~\cite{CT10} PDF sets at the corresponding order in perturbation theory (NNLO or NLO) are used in all the predictions above. The associated uncertainties are at the level of 2\%. They are obtained by reweighting the samples to the various eigenvectors of the CT10 set and scaling the resulting uncertainty to 68\% confidence level. Non-perturbative effects are evaluated by comparing \textsc{Sherpa} signal samples with or without hadronization and the underlying event, and applied as corrections to the calculations. The impact on the differential cross sections is typically below 5\%, reaching as high as 10\% in some regions. The full size of the effect is added as a systematic uncertainty. The dominant uncertainties arise from the choices of renormalization and factorization scales, set by default to \mgg. They are estimated simultaneously by varying the nominal values by a factor of two up and down and taking the largest differences with respect to the nominal predictions in each direction as uncertainties. The \textsc{Diphox} computation also requires the definition of a fragmentation scale, which is varied together with the factorization scale for the uncertainty estimation. The total uncertainties are $^{+15}_{-10}$\% and $^{+9}_{-6}$\% in the inclusive case and within $\pm30\%$ and $\pm20\%$ in almost all bins for \textsc{Diphox} and \textsc{2$\gamma$NNLO}, respectively. Only the central values of the predictions were provided by the authors of \textsc{Resbos}.

A recent implementation of \textsc{Sherpa}, version 2.2.1~\cite{Siegert:2016bre}, consistently combines  parton-level calculations of varying jet multiplicity up to NLO\footnote{The $\gamma\gamma$ and $\gamma\gamma$ + 1 parton processes are generated at NLO accuracy, while the $\gamma\gamma$ + 2 partons and $\gamma\gamma$ + 3 partons processes are generated at LO. Charm and bottom quarks are included in these matrix elements in the massless approximation.}~\cite{Krauss:2001iv, Cascioli:2011va} with parton showering~\cite{Schumann:2007mg} while avoiding double-counting effects~\cite{Hoeche:2012yf}. The NNPDF3.0 NNLO PDFs~\cite{Ball:2014uwa} are used in conjunction with the corresponding \textsc{Sherpa} default tuning. Dynamic factorization and renormalization scales are adopted~\cite{Siegert:2016bre}. The associated uncertainties are estimated by varying each by a factor of two up and down and retaining the largest variations. They range from 10\% to 40\% across the various regions, with typical values around 20\%. The relative uncertainty in the integrated cross section is $^{+19}_{-13}$\%.

\begin{figure}[!ht]
\centering
\includegraphics[width=0.8\textwidth]{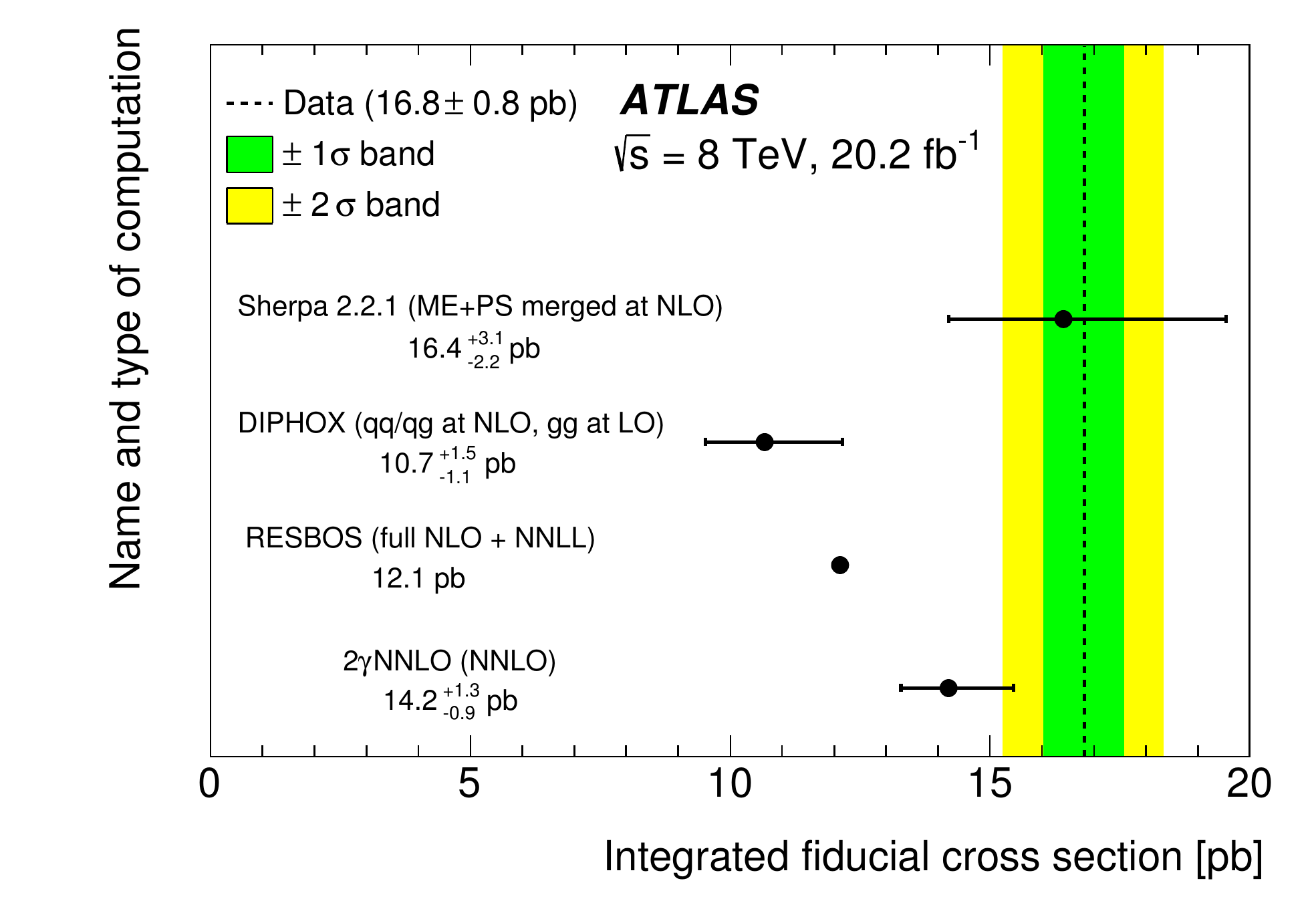}
\caption{Measured fiducial cross section compared to the predictions from \textsc{Sherpa} 2.2.1, \textsc{Diphox}, \textsc{Resbos} and \textsc{2$\gamma$NNLO}. The estimation of the uncertainties in the theoretical predictions are described in the text. Only the central value is shown for \textsc{Resbos}. The green (yellow) band represents the one- (two-)standard deviation uncertainty, including both the statistical and systematic uncertainties in the measurement added in quadrature.}
\label{fig:int_cross_section_vs_theory}
\end{figure}

Figure~\ref{fig:int_cross_section_vs_theory} shows the comparison of the integrated fiducial cross-section measurement with the calculations described above. The NLO prediction from \textsc{Diphox} is 36\% lower than the measured value, which corresponds to more than two standard deviations of both the experimental and theoretical uncertainties. The cross sections calculated with \textsc{Resbos} and \textsc{2$\gamma$NNLO} underestimate the experimental result by 28\% and 16\%, respectively. The prediction from \textsc{Sherpa} 2.2.1 is in agreement with the data.

\begin{figure}[htbp]
\centering
\includegraphics[width=0.4\textwidth]{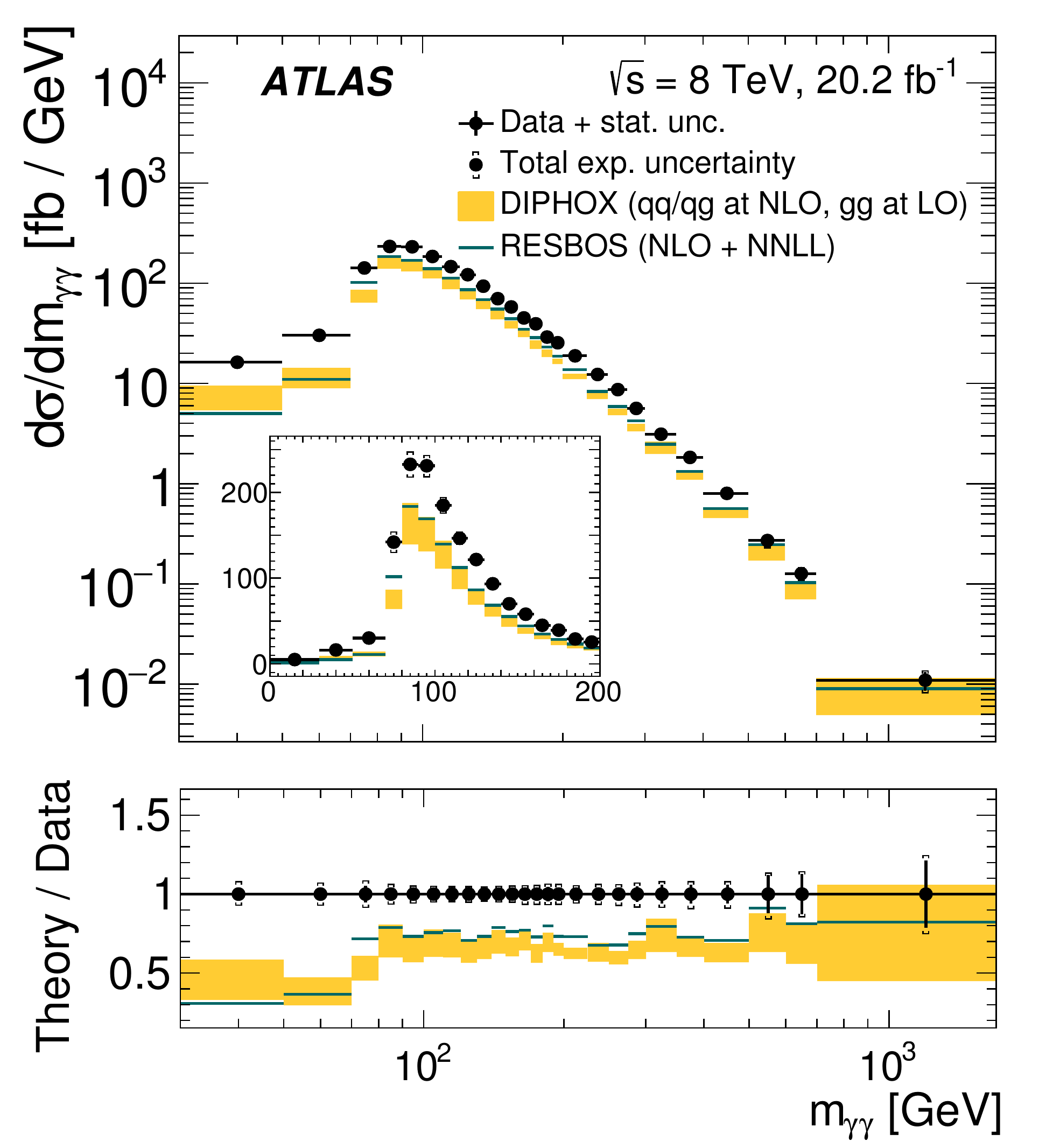}
\includegraphics[width=0.4\textwidth]{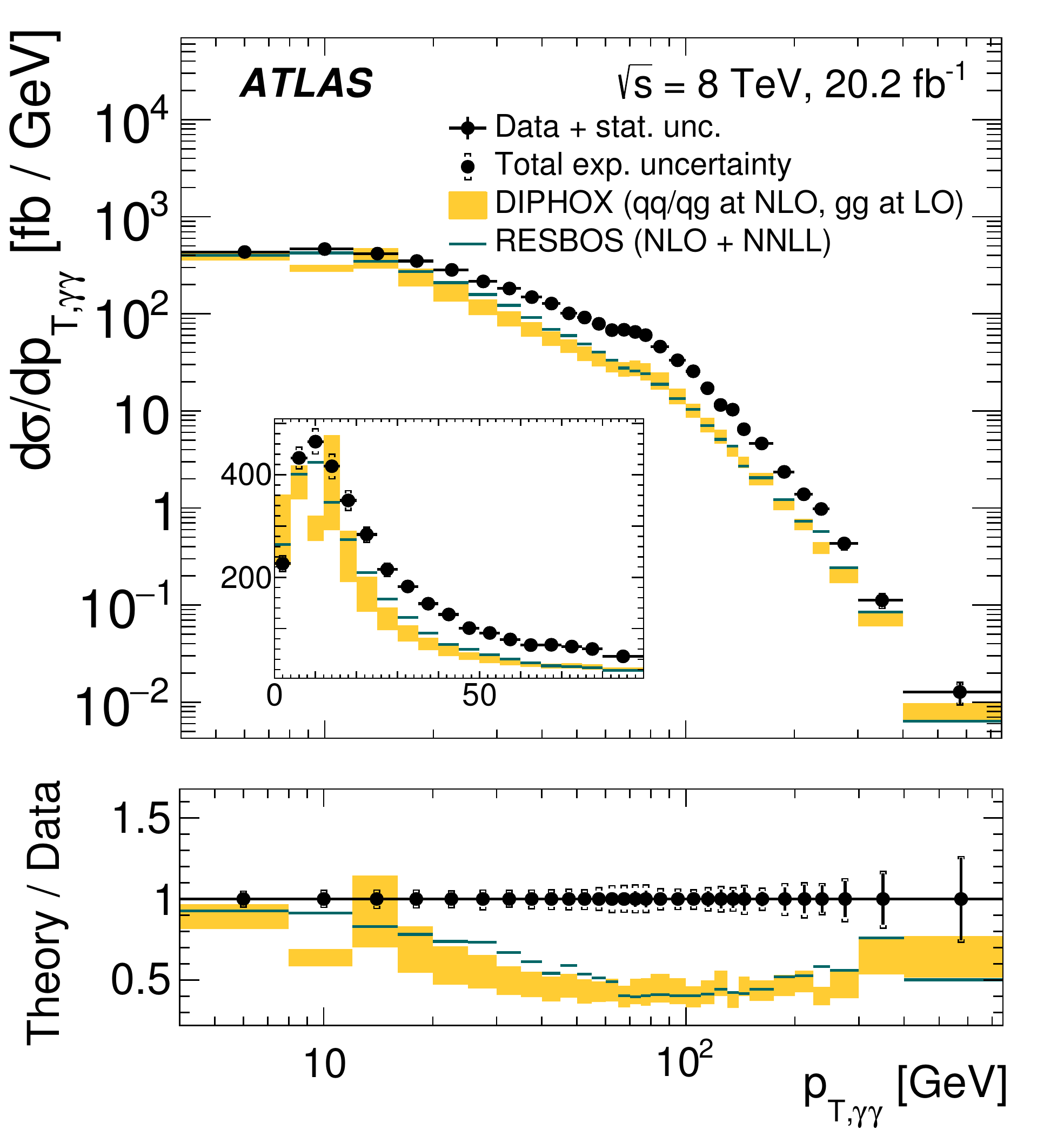}
\includegraphics[width=0.4\textwidth]{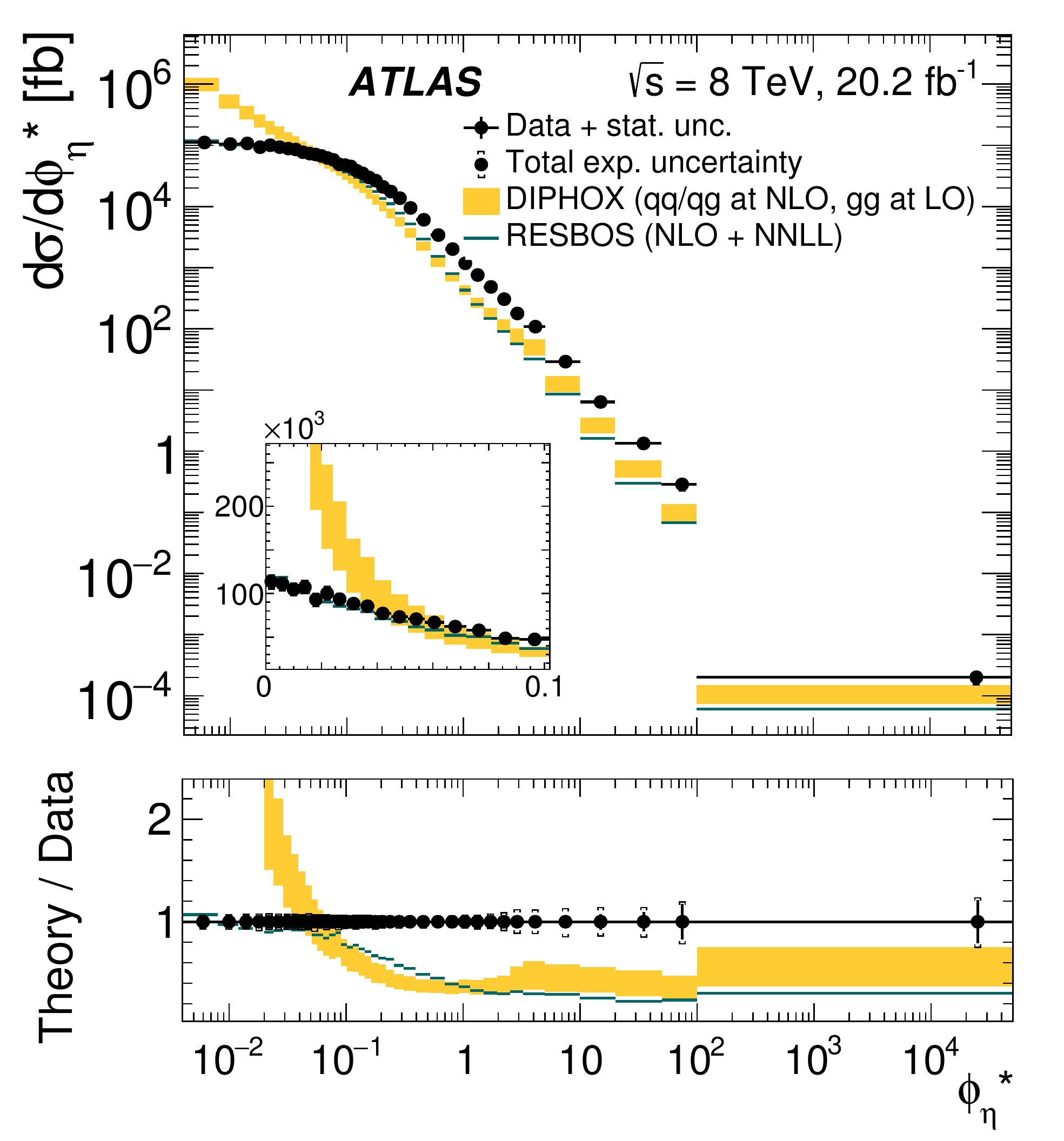}
\includegraphics[width=0.4\textwidth]{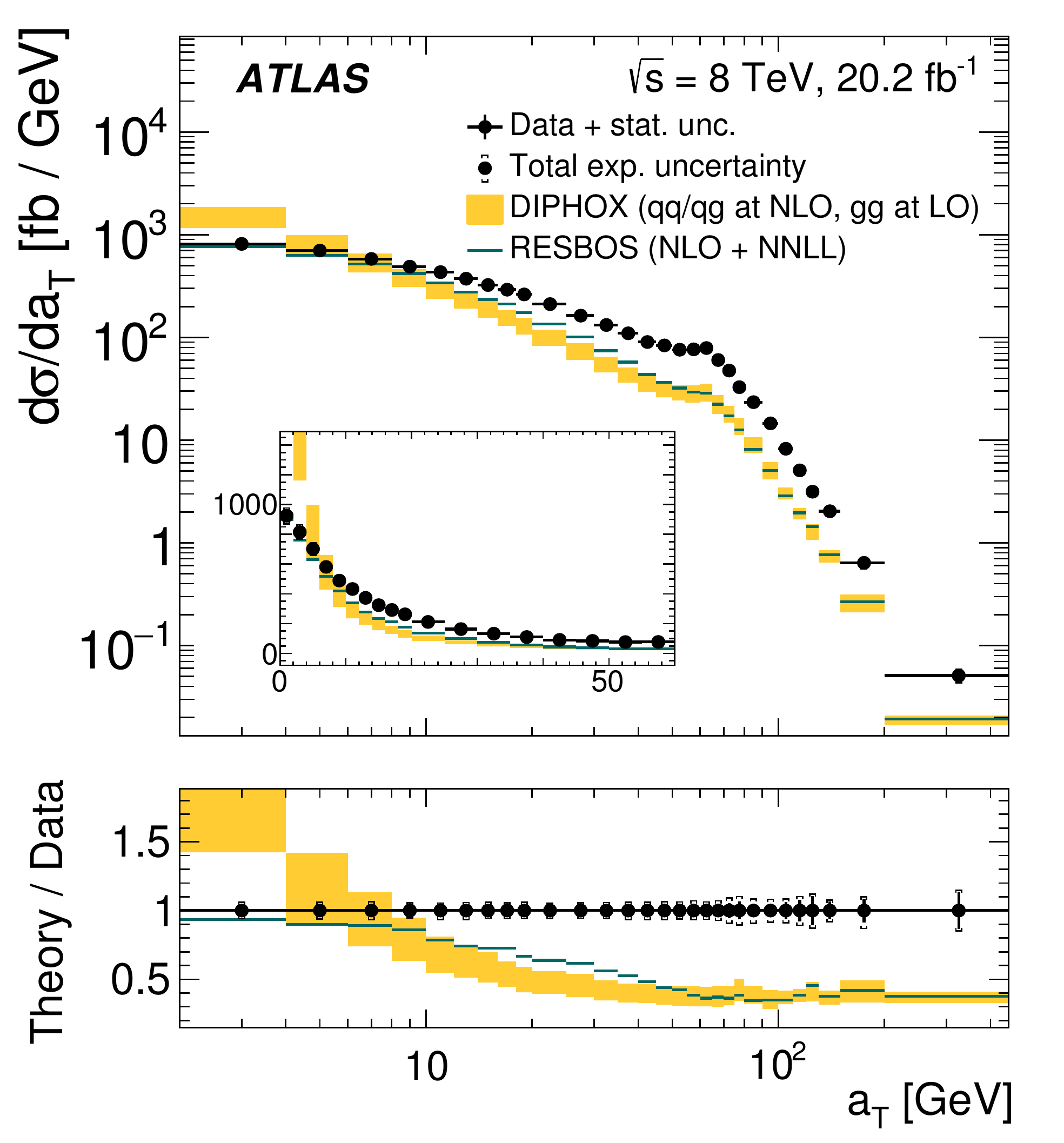}
\includegraphics[width=0.4\textwidth]{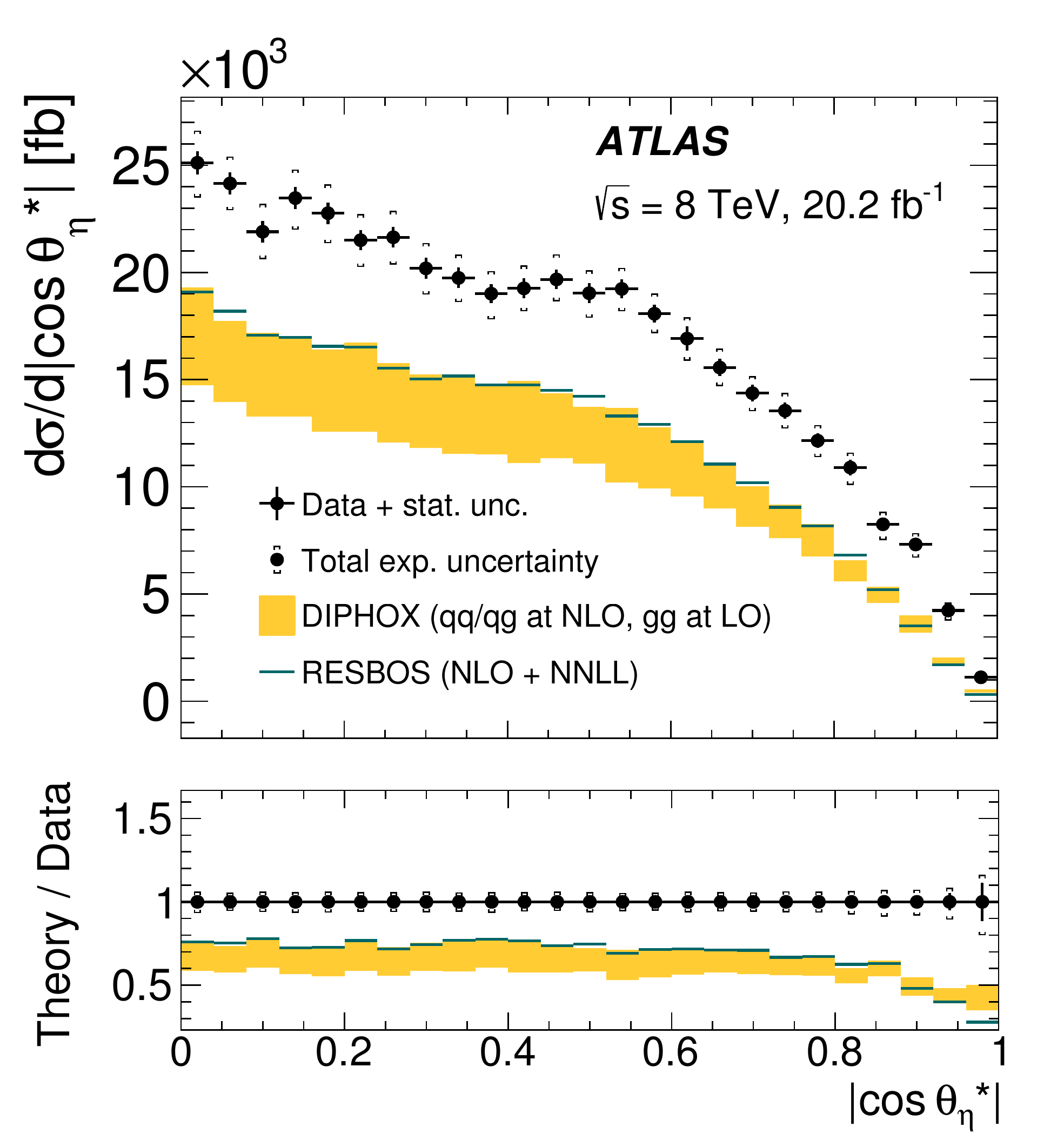}
\includegraphics[width=0.4\textwidth]{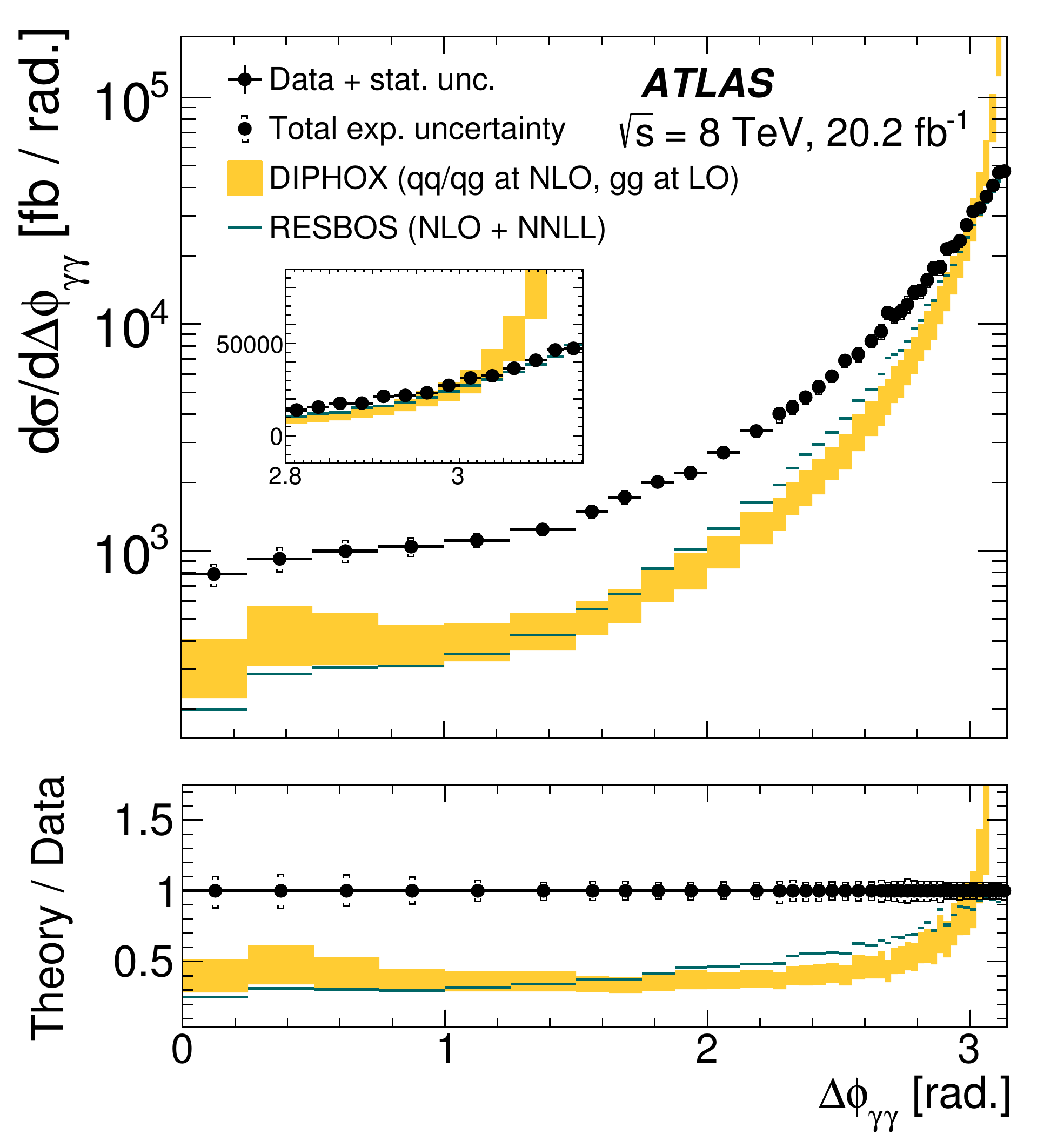}
\caption{Differential cross sections as functions of the various observables compared to the predictions from \textsc{Diphox} and \textsc{Resbos}. At the bottom of each plot, the ratio of the prediction to the data is shown. The bars and bands around the data and theoretical predictions represent the statistical and systematic uncertainties, estimated as described in the text. Only the central values are shown for \textsc{Resbos}. Negative cross-section values are obtained with \textsc{Diphox} in the first (last) bin of \at\ and \phistareta\ (\deltaphi) and therefore are not shown (see text).}
\label{fig:cross_sections_vs_NLO}
\end{figure}

\begin{figure}[htbp]
\centering
\includegraphics[width=0.4\textwidth]{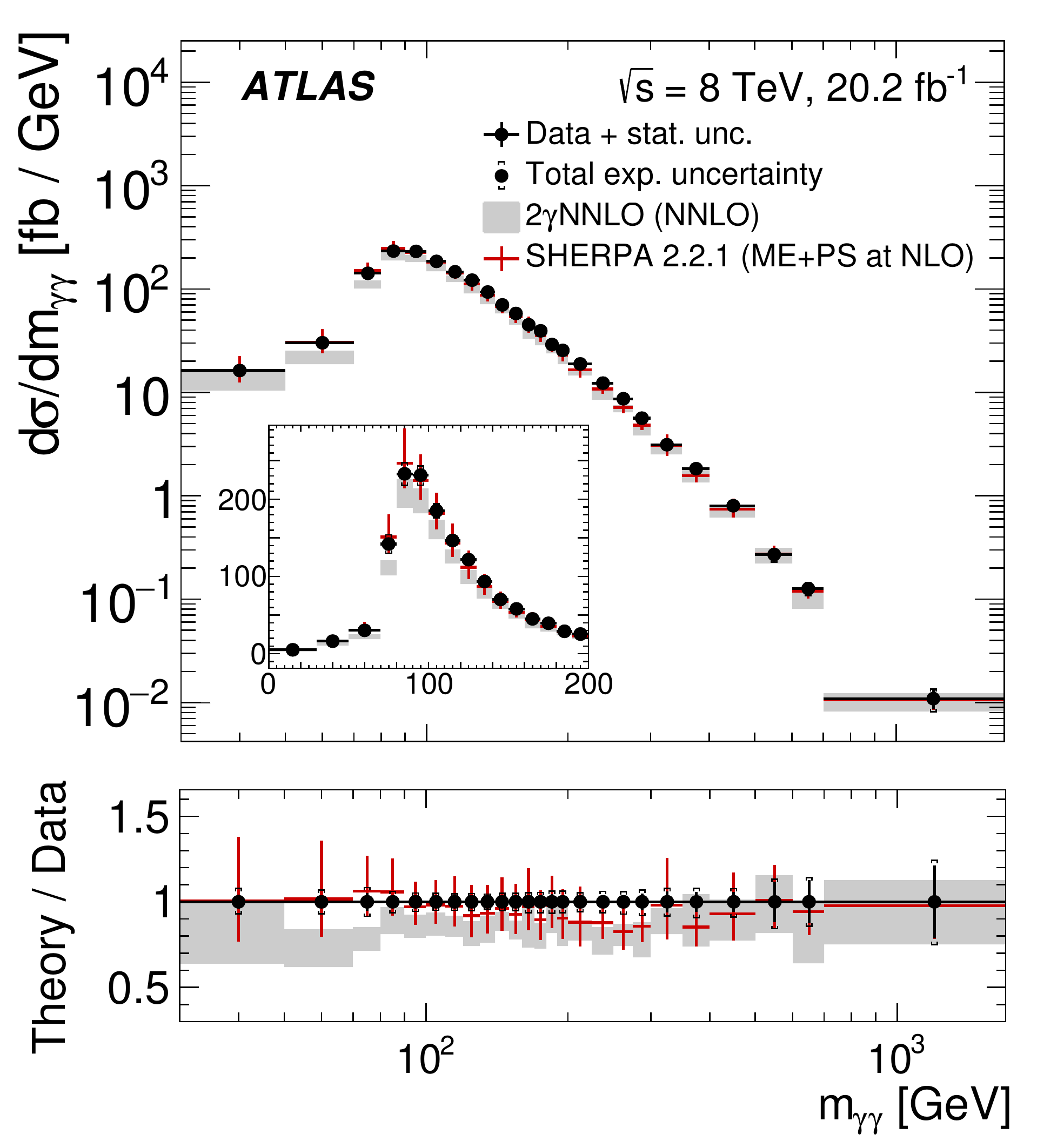}
\includegraphics[width=0.4\textwidth]{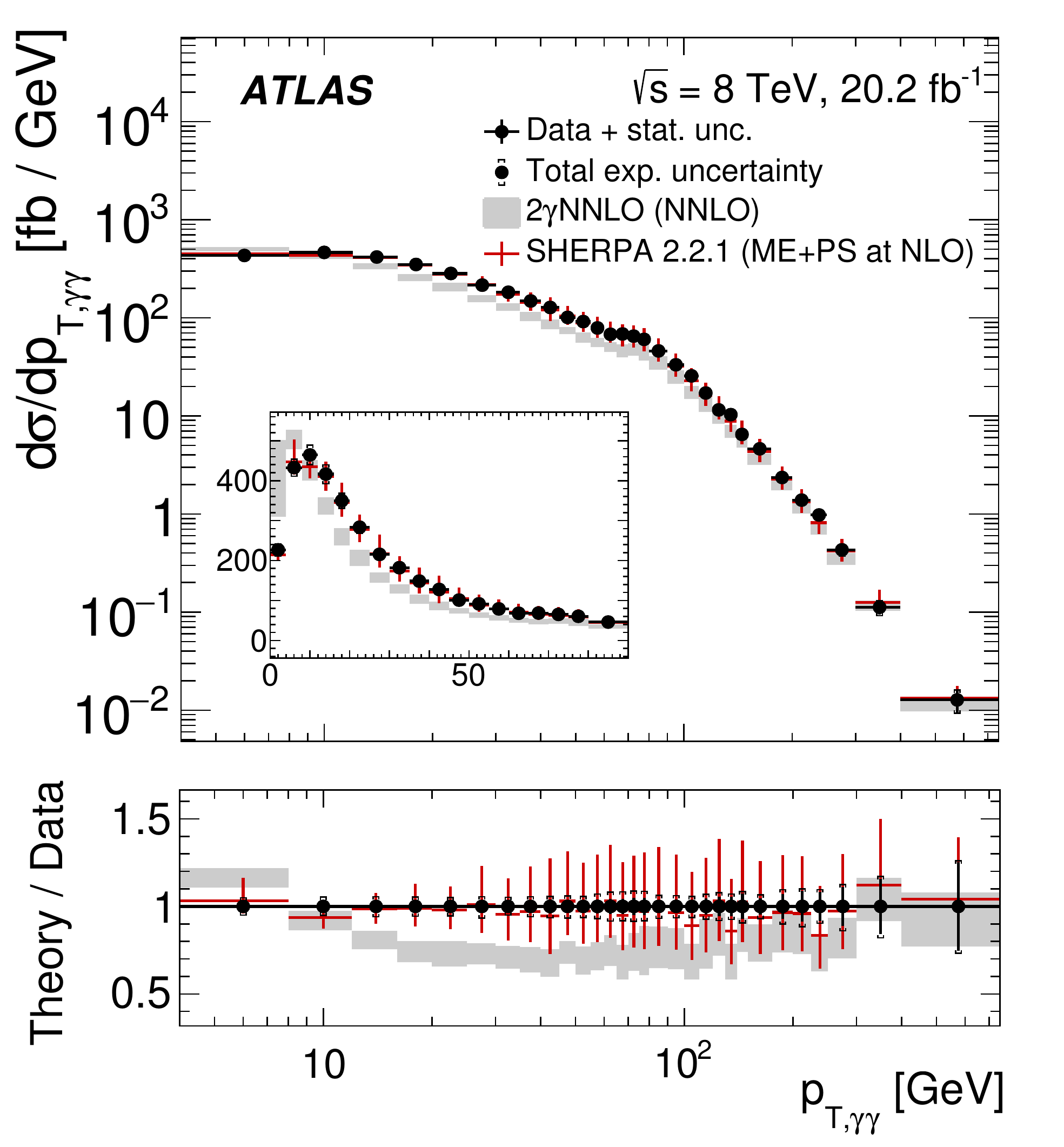}
\includegraphics[width=0.4\textwidth]{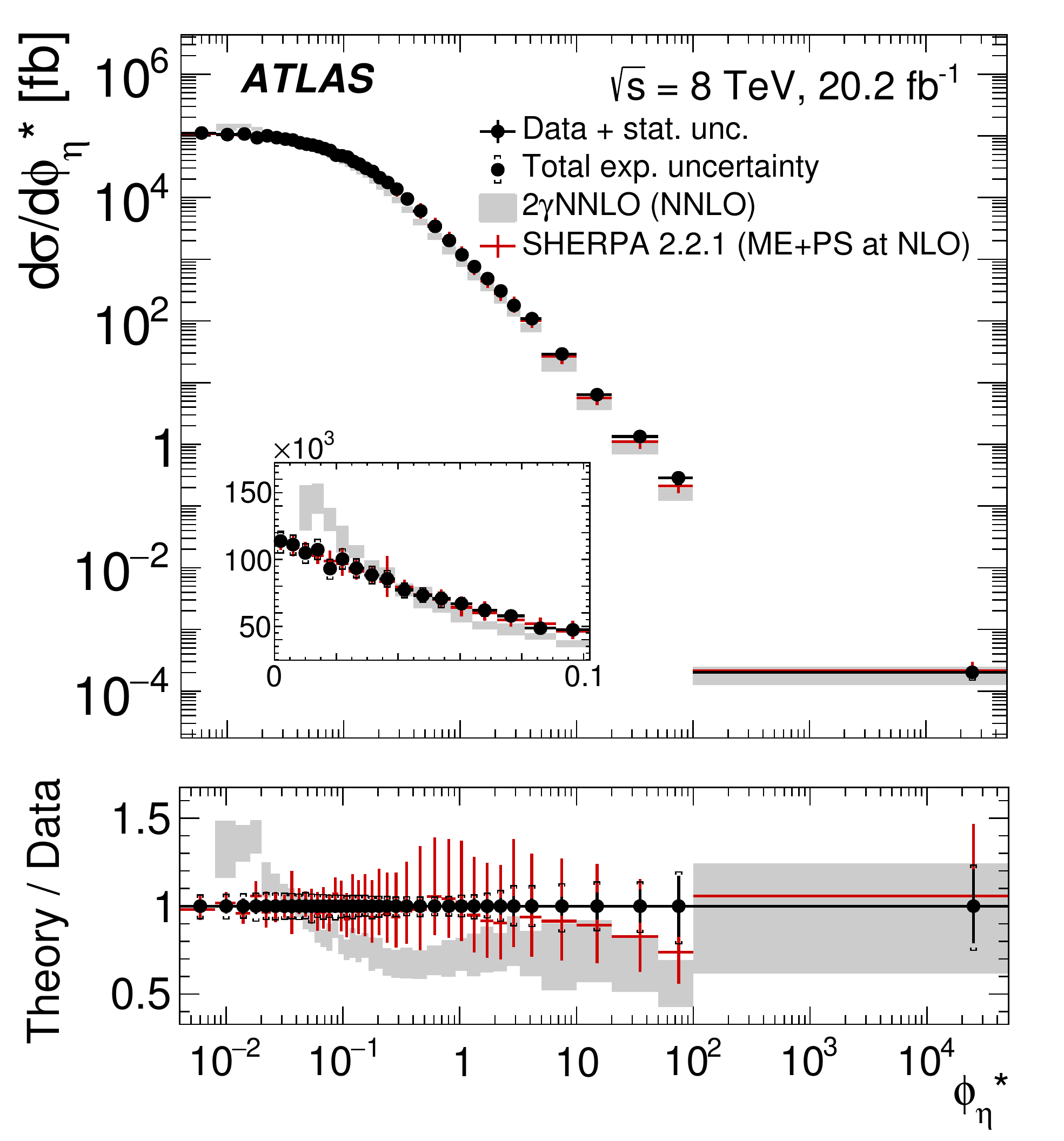}
\includegraphics[width=0.4\textwidth]{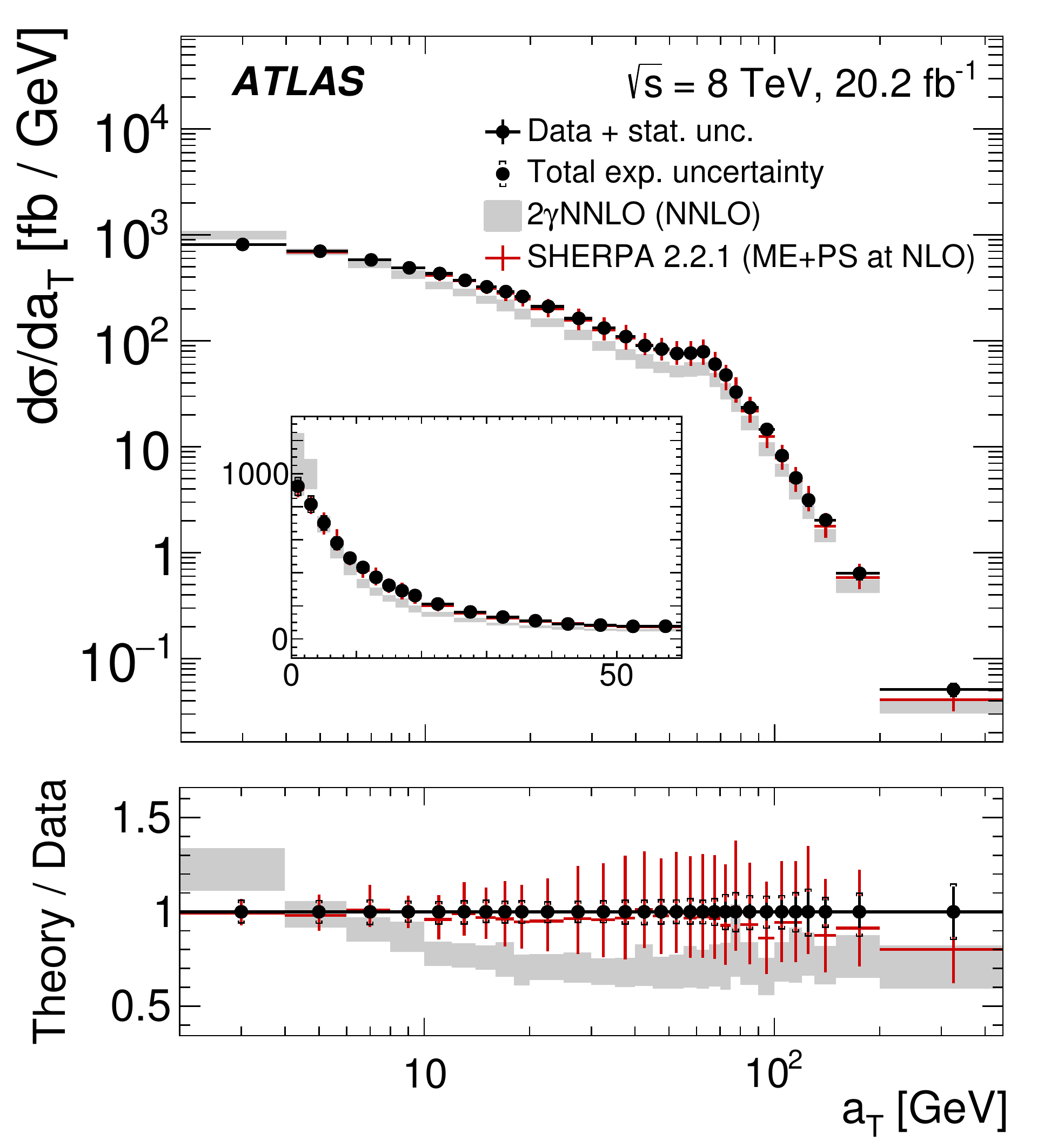}
\includegraphics[width=0.4\textwidth]{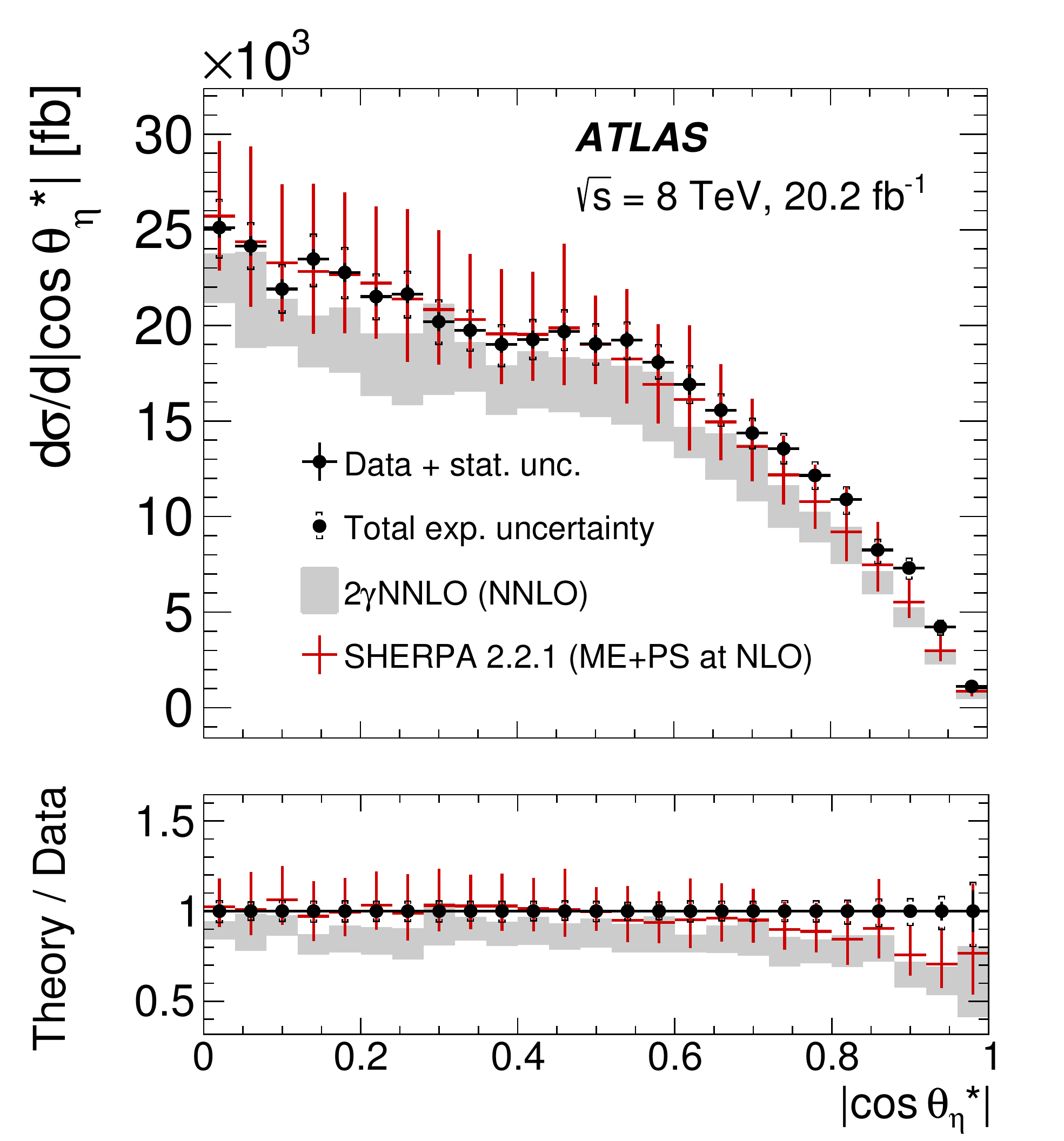}
\includegraphics[width=0.4\textwidth]{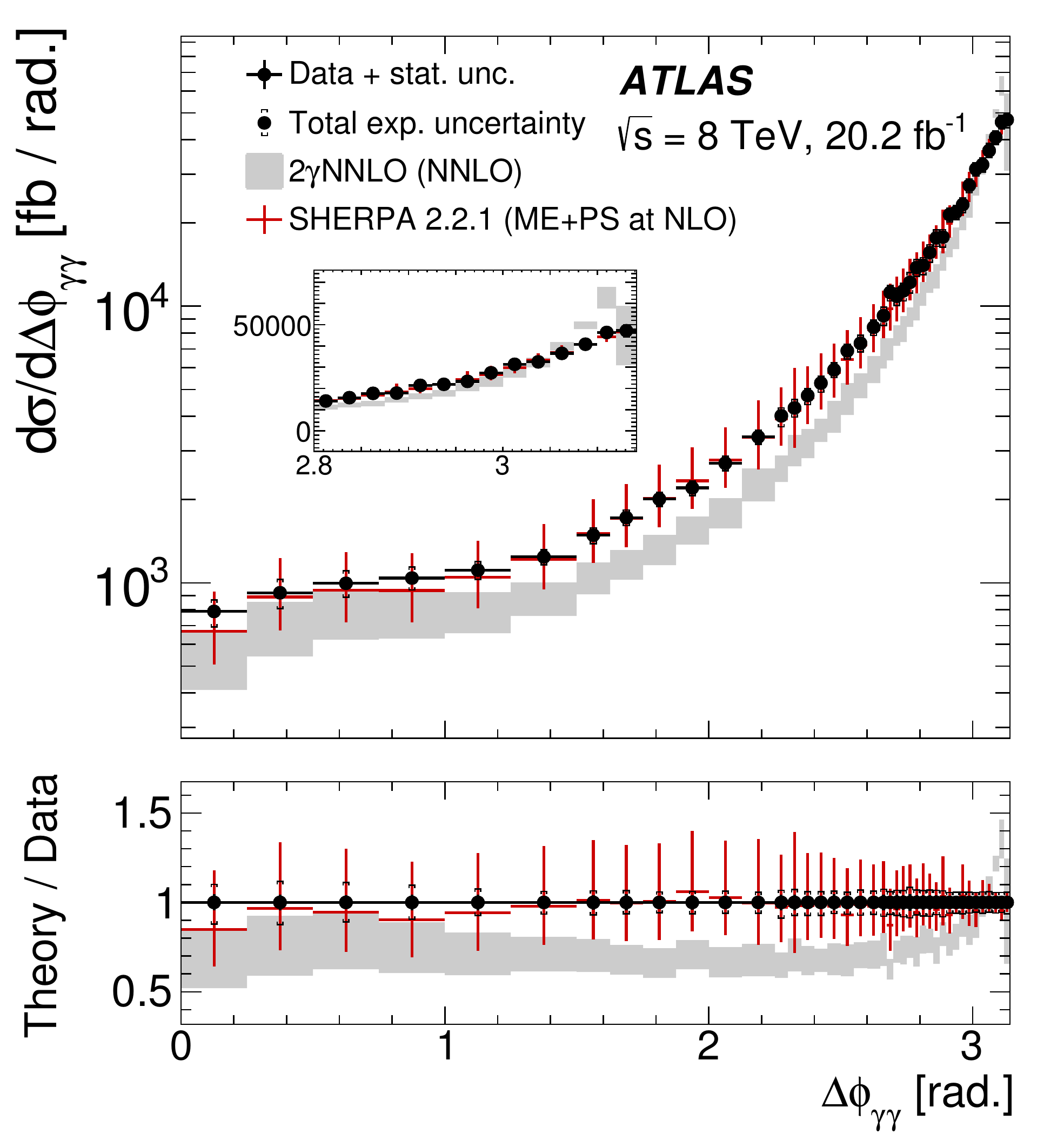}
\caption{Differential cross sections as functions of the various observables compared to the predictions from \textsc{Sherpa}~2.2.1 and \textsc{2$\gamma$NNLO}. At the bottom of each plot, the ratio of the prediction to the data is shown. The bars and bands around the data and theoretical predictions represent the statistical and systematic uncertainties, estimated as described in the text. Negative cross-section values are obtained with \textsc{2$\gamma$NNLO} when varying the renormalisation scale in the first two bins of \phistareta\ and therefore are not shown (see text).}
\label{fig:cross_sections_vs_SherpaNLO_2gNNLO}
\end{figure}

The differential cross-section measurements are compared to the predictions in Figures~\ref{fig:cross_sections_vs_NLO} and \ref{fig:cross_sections_vs_SherpaNLO_2gNNLO}. Again, the predictions from \textsc{Sherpa} 2.2.1 are in agreement with the data for the differential distributions. The cross sections at large values of \acosthetastareta\ and values of \mgg\ between 100~\GeV\ and 500~\GeV\ are slightly underestimated, although compatible within errors. The inclusion of soft-gluon resummation (\textsc{Resbos}) or a parton shower (\textsc{Sherpa} 2.2.1) provides a correct description of regions of the phase space that are sensitive to infrared emissions, i.e. at low values of \ptgg, \at\ and \phistareta\ or  $\deltaphi \sim \pi$. Fixed-order calculations, instead, are not expected to give reliable predictions in these regions. Negative cross-section values are obtained with \textsc{Diphox} and \textsc{2$\gamma$NNLO} in some cases. The turn-on behavior observed in \mgg\ and located at about the sum of the minimum \et\ required for each photon is fairly well reproduced by all the predictions. The `shoulders'  observed in \ptgg\ and \at, induced by the requirements on the photon kinematics combined with radiative effects~\cite{Binoth:2000zt}, are also well reproduced. In other regions of the phase space, disagreements of up to a factor of two between the parton-level computations at NLO and the measurements are observed. The inclusion of NNLO corrections is not sufficient to reproduce the measurements.

 \FloatBarrier

\section{Conclusion}
\label{sec:conclusion}

Measurements of the production cross section of two isolated photons in proton--proton collisions at a center-of-mass energy of $\sqrt{s}=8~\TeV$ have been presented. 
The results use a data set of an integrated luminosity of 20.2~fb$^{-1}$ recorded by the ATLAS detector at the LHC in 2012. Events are considered in which the two photons with the highest transverse energies in the event are within the acceptance of the calorimeter ($|\eta^{\gamma}|<1.37$ or $1.56<|\eta^{\gamma}|<2.37$), have transverse energies greater than 40~\GeV\ and 30~\GeV, respectively, and an angular separation of $\Delta R_{\gamma\gamma} > 0.4$. The measurements are unfolded to particle level imposing a maximum transverse isolation energy of 11~\GeV\ within a cone of $\Delta R = 0.4$ for both photons. 

The measured integrated cross section within the fiducial volume is $16.8 \pm 0.8$~pb. The uncertainties are dominated by systematic effects that have been reduced compared to the previous ATLAS measurement at $\sqrt{s}=7~\TeV$ due to an improved method to estimate the background and improved corrections to the modeling of the calorimeter isolation variable in simulated samples. Predicted cross sections from fixed-order QCD calculations implemented in \textsc{Diphox} and \textsc{Resbos} at next-to-leading order, and in \textsc{2$\gamma$NNLO} at next-to-next-to-leading order, are about 36\%, 28\% and 16\% lower than the data, respectively. The relative errors associated to the predictions from \textsc{Diphox} (\textsc{2$\gamma$NNLO}) are 10--15\% (5--10\%).

Differential cross sections are measured as functions of six observables -- the diphoton invariant mass, the absolute value of the cosine of the scattering angle with respect to the direction of the proton beams, the opening angle between the photons in the azimuthal plane, the diphoton transverse momentum and two related variables (\at\ and \phistareta) -- with uncertainties typically below 5\% per bin, reaching as high as 25\% in a few bins with low numbers of data events. The effects of infrared emissions, probed precisely by measuring the cross section as functions of \at\ and \phistareta, are well reproduced by the inclusion of soft-gluon resummation at the next-to-next-to-leading-logarithm accuracy. However, in most parts of the phase space, the predictions above are unable to reproduce the data. The discrepancies can reach a factor of two in many regions, beyond the theoretical uncertainties, which are typically below 20\%.

The predictions of a parton-level calculation of varying jet multiplicity up to NLO matched to a parton-shower algorithm in \textsc{Sherpa} 2.2.1~provide an improved description of the data compared to all the other computations considered in this paper and are in good agreement with the measurements, for both the integrated and differential cross sections.

\section*{Acknowledgements}


We thank CERN for the very successful operation of the LHC, as well as the
support staff from our institutions without whom ATLAS could not be
operated efficiently.

We acknowledge the support of ANPCyT, Argentina; YerPhI, Armenia; ARC, Australia; BMWFW and FWF, Austria; ANAS, Azerbaijan; SSTC, Belarus; CNPq and FAPESP, Brazil; NSERC, NRC and CFI, Canada; CERN; CONICYT, Chile; CAS, MOST and NSFC, China; COLCIENCIAS, Colombia; MSMT CR, MPO CR and VSC CR, Czech Republic; DNRF and DNSRC, Denmark; IN2P3-CNRS, CEA-DSM/IRFU, France; SRNSF, Georgia; BMBF, HGF, and MPG, Germany; GSRT, Greece; RGC, Hong Kong SAR, China; ISF, I-CORE and Benoziyo Center, Israel; INFN, Italy; MEXT and JSPS, Japan; CNRST, Morocco; NWO, Netherlands; RCN, Norway; MNiSW and NCN, Poland; FCT, Portugal; MNE/IFA, Romania; MES of Russia and NRC KI, Russian Federation; JINR; MESTD, Serbia; MSSR, Slovakia; ARRS and MIZ\v{S}, Slovenia; DST/NRF, South Africa; MINECO, Spain; SRC and Wallenberg Foundation, Sweden; SERI, SNSF and Cantons of Bern and Geneva, Switzerland; MOST, Taiwan; TAEK, Turkey; STFC, United Kingdom; DOE and NSF, United States of America. In addition, individual groups and members have received support from BCKDF, the Canada Council, CANARIE, CRC, Compute Canada, FQRNT, and the Ontario Innovation Trust, Canada; EPLANET, ERC, ERDF, FP7, Horizon 2020 and Marie Sk{\l}odowska-Curie Actions, European Union; Investissements d'Avenir Labex and Idex, ANR, R{\'e}gion Auvergne and Fondation Partager le Savoir, France; DFG and AvH Foundation, Germany; Herakleitos, Thales and Aristeia programmes co-financed by EU-ESF and the Greek NSRF; BSF, GIF and Minerva, Israel; BRF, Norway; CERCA Programme Generalitat de Catalunya, Generalitat Valenciana, Spain; the Royal Society and Leverhulme Trust, United Kingdom.

The crucial computing support from all WLCG partners is acknowledged gratefully, in particular from CERN, the ATLAS Tier-1 facilities at TRIUMF (Canada), NDGF (Denmark, Norway, Sweden), CC-IN2P3 (France), KIT/GridKA (Germany), INFN-CNAF (Italy), NL-T1 (Netherlands), PIC (Spain), ASGC (Taiwan), RAL (UK) and BNL (USA), the Tier-2 facilities worldwide and large non-WLCG resource providers. Major contributors of computing resources are listed in Ref.~\cite{ATL-GEN-PUB-2016-002}.

\printbibliography

\newpage 
\begin{flushleft}
{\Large The ATLAS Collaboration}

\bigskip

M.~Aaboud$^\textrm{\scriptsize 137d}$,
G.~Aad$^\textrm{\scriptsize 88}$,
B.~Abbott$^\textrm{\scriptsize 115}$,
J.~Abdallah$^\textrm{\scriptsize 8}$,
O.~Abdinov$^\textrm{\scriptsize 12}$$^{,*}$,
B.~Abeloos$^\textrm{\scriptsize 119}$,
S.H.~Abidi$^\textrm{\scriptsize 161}$,
O.S.~AbouZeid$^\textrm{\scriptsize 139}$,
N.L.~Abraham$^\textrm{\scriptsize 151}$,
H.~Abramowicz$^\textrm{\scriptsize 155}$
% ...
\bigskip
\\
% ...
$^{8}$ Department of Physics, The University of Texas at Arlington, Arlington TX, United States of America\\
% ...
$^{12}$ Institute of Physics, Azerbaijan Academy of Sciences, Baku, Azerbaijan\\
% ...
$^{88}$ CPPM, Aix-Marseille Universit{\'e} and CNRS/IN2P3, Marseille, France\\
% ...
$^{115}$ Homer L. Dodge Department of Physics and Astronomy, University of Oklahoma, Norman OK, United States of America\\
% ...
$^{119}$ LAL, Univ. Paris-Sud, CNRS/IN2P3, Universit{\'e} Paris-Saclay, Orsay, France\\
% ...
$^{137}$ $^{(a)}$ Facult{\'e} des Sciences Ain Chock, R{\'e}seau Universitaire de Physique des Hautes Energies - Universit{\'e} Hassan II, Casablanca; $^{(b)}$ Centre National de l'Energie des Sciences Techniques Nucleaires, Rabat; $^{(c)}$ Facult{\'e} des Sciences Semlalia, Universit{\'e} Cadi Ayyad, LPHEA-Marrakech; $^{(d)}$ Facult{\'e} des Sciences, Universit{\'e} Mohamed Premier and LPTPM, Oujda; $^{(e)}$ Facult{\'e} des sciences, Universit{\'e} Mohammed V, Rabat, Morocco\\
% ...
$^{139}$ Santa Cruz Institute for Particle Physics, University of California Santa Cruz, Santa Cruz CA, United States of America\\
% ...
$^{151}$ Department of Physics and Astronomy, University of Sussex, Brighton, United Kingdom\\
% ...
$^{155}$ Raymond and Beverly Sackler School of Physics and Astronomy, Tel Aviv University, Tel Aviv, Israel\\
% ...
$^{161}$ Department of Physics, University of Toronto, Toronto ON, Canada\\
% ...
$^{a}$ Also at Department of Physics, King's College London, London, United Kingdom\\
$^{b}$ Also at Institute of Physics, Azerbaijan Academy of Sciences, Baku, Azerbaijan\\
$^{c}$ Also at Novosibirsk State University, Novosibirsk, Russia\\
$^{d}$ Also at TRIUMF, Vancouver BC, Canada\\
$^{e}$ Also at Department of Physics {\&} Astronomy, University of Louisville, Louisville, KY, United States of America\\
% ...
$^{*}$ Deceased
\end{flushleft}


\end{document}